\documentclass{article}

\usepackage[utf8]{inputenc}
\usepackage{hyperref}       
\usepackage{graphicx}
\usepackage[numbers]{natbib}
\usepackage{amsmath}
\usepackage{amssymb}
\usepackage{subcaption}
\usepackage{multirow}
\usepackage{booktabs}       
\usepackage[a4paper, total={6in, 8in}]{geometry}

\title{\huge Improving Parametric Neural Networks \\ for High-Energy Physics (and Beyond)\footnote{Published as a journal paper at Machine Learning: Science and Technology. Available at \url{https://doi.org/10.1088/2632-2153/ac917c}.}}

\author{
    \href{https://orcid.org/0000-0002-0399-8836}{\includegraphics[scale=0.06]{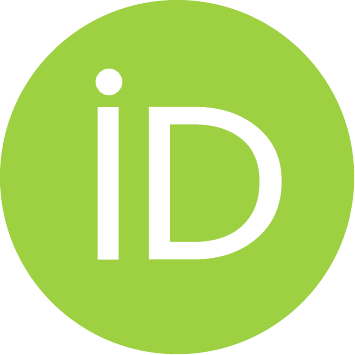}\hspace{1mm}Luca ~Anzalone}$^{1,3*}$,
    \href{https://orcid.org/0000-0003-0780-8785}{\includegraphics[scale=0.06]{orcid.pdf} Tommaso ~Diotalevi}$^{1,2,3}$,
    \href{https://orcid.org/0000-0002-0835-9574}{\includegraphics[scale=0.06]{orcid.pdf} Daniele ~Bonacorsi}$^{1,3}$
\\
	\\
	$^{1}$\small{Department of Physics and Astronomy (DIFA), University of Bologna, Bologna, Italy} \\
	$^{2}$\small{European Organization for Nuclear Research (CERN), Geneva, Switzerland}\\
    $^{3}$\small{Istituto Nazionale di Fisica Nucleare (INFN), Sezione di Bologna, Italy}\\
    $^{*}$\small{Corresponding author, email at \texttt{luca.anzalone2@unibo.it}}
}

\date{}

\hypersetup{
    pdftitle={Improving Parametric Neural Networks for High-Energy Physics (and Beyond)},
    pdfauthor={Luca ~Anzalone, Tommaso ~Diotalevi, Daniele ~Bonacorsi},
    pdfkeywords={Parametric Neural Networks, Conditioning Mechanisms, Signal-background Classification, High-Energy Physics},
}

\newcommand{\gev}{GeV}%
\newcommand{\mass}{\mathcal{M}}%
\newcommand{\hepmass}{\texttt{HEPMASS}}%
\newcommand{\hepimb}{\texttt{HEPMASS-IMB}}%

\begin{document}

\maketitle

\begin{abstract}
	\noindent Signal-background classification is a central problem in High-Energy Physics (HEP), that plays a major role for the discovery of new fundamental particles. A recent method -- the Parametric Neural Network (pNN) -- leverages multiple signal mass hypotheses as an additional input feature to effectively replace a whole set of individual classifiers, each providing (in principle) the best response for the corresponding mass hypothesis. In this work we aim at deepening the understanding of pNNs in light of real-world usage. We discovered several peculiarities of parametric networks, providing intuition, metrics, and guidelines to them. We further propose an alternative parametrization scheme, resulting in a new parametrized neural network architecture: the AffinePNN; along with many other generally applicable improvements, like the \textit{balanced training} procedure. Finally, we extensively and empirically evaluate our models on the \hepmass\ dataset, along its \textit{imbalanced} version (called \hepimb) we provide here for the first time, to further validate our approach. Provided results are in terms of the impact of the proposed design decisions, classification performance, and interpolation capability, as well.
\end{abstract}

\section{Introduction}
\label{sec:intro}
Selecting events that contain interesting processes is a fundamental requirement of HEP experiments and one of the most established areas in advanced computing techniques i.e. Machine and Deep Learning \cite{goodfellow2016deep}. Physicists are interested in rare events (yield by the collision of known particles), following their theoretical assumptions, measuring the fraction of events that contain a specific decay channel. These rare events -- the so-called \textit{signal} -- must be separated out from the \textit{background}, i.e. anything else originating from already known processes. The usual way of doing that relies on building an event selection algorithm, estimating its efficiency on selecting signal and rejecting background, and measuring the count of events passing it. Being able to effectively separate background events from the signal is a central problem in HEP, that can help the discovery of new fundamental particles with further analysis. 
Compared to traditional, expert-designed algorithms based on single cuts driven by physical considerations, Machine Learning algorithms, e.g. (boosted) decision trees (DTs) \cite{friedman2001elements, chatrchyan2012observation} and neural networks (NNs) \cite{goodfellow2016deep, baldi2014searching}, have two main advantages: first of all, they are usually able to deliver an higher selection efficiency; secondly, they save effort, by replacing an HEP-specific manual algorithm solution with an application of a general method, generally stolen from an AI research and adopted also in other fields of study. Since, prior to the analysis, a signal event is not known, these algorithms are trained on synthetic data obtained through expensive Monte-Carlo simulations, trying to mimic data as coming from real-world collisions of particles in accelerators, like the CERN Large Hadron Collider (LHC) \cite{Evans_2008}. \\ 
\\
In addition to the above mentioned challenges, some searches do not even have a clear theoretical prediction on the exact mass values for such potential new particles: for instance, several new physics analyses prefer to keep such mass values floating, generating $M$ different Monte-Carlo samples at different mass hypotheses and splitting the analysis in $M$ different searches, each one with a fixed mass value \cite{cms2019qli, cms2017search}. \\
\\
Before the introduction of parametrized neural networks (pNNs) \cite{baldi2016parameterized} (described in section \ref{subsec:pnn}), therefore, the canonical approach for signal-background classification consisted of training a set of \textit{individual and independent classifiers} \cite{baldi2014searching}, each of them trained on a specific signal mass hypothesis $m_i\in \mathcal{M}$ (thus, on a subset of the available data, somehow). The benefit of individual classifiers is that each of them should maximize the separation of the two classes for a single mass hypothesis. The major drawback is that researchers have to design, train, tune, evaluate, and maintain $|\mathcal M|$ of such classifiers, one for each mass\footnote{In this work, the term \textit{mass} refers to the additional input of the parametric network, $m$: also called \textit{mass feature} by Baldi et al \cite{baldi2016parameterized}. By $m_i$ we denote a generic value of the signal generating mass (or signal mass hypotheses), whose set of values is represented by $\mass$. In general, we don't use the term mass to imply the reconstructed (or invariant) mass of the considered particle decay, but at most its theoretical parameter, $m_X$.} 
($m_i \in \mathcal M$), often with different hyperparameters, whose amount can easily be in the order of tens: this can become easily unfeasible. The other issue is about classification performance. Specifically talking about neural networks, they greatly benefit from the \textit{sharing of weights} and \textit{distributed representation} \cite{goodfellow2016deep} for improved accuracy, reduced training time, and thus increased data efficiency. Parametric neural networks aim at mitigating those issues by means of an additional input (the signal's generating mass), while also bringing additional benefits for the sake of more effective research.

The overall contributions of our work are summarized as follows:
\begin{itemize}
    \item We propose a novel and more challenging benchmark dataset called \hepimb\ \cite{anzalone2022hepimb} (section \ref{subsec:hep_imb}), to overcome the simplicity of \hepmass\  \cite{baldi2015hepmass}, trying to emulate real-world scenarios more closely.
    \item We developed a novel parametrization scheme: the \textit{affine parametric neural network} (section \ref{subsec:affine_arch}).
    \item We describe and empirically study several design decisions for building effective pNNs (section \ref{sec:design_choices}). In this regard, we address the issue of how to assign (or distribute) the mass feature for the background events (section \ref{subsec:bkg}), as well as providing a novel \textit{balanced training} procedure (\ref{subsec:balanced_training}).
    \item We attempt a first characterization of the \textit{properties} that parametric networks have (section \ref{sec:properties}).
    \item Regarding interpolation (section \ref{subsec:interpolation}), we study it in depth: showing a failure case, and providing guidelines for its assessment.
\end{itemize}

\section{Related Work}
\label{sec:rel_work}
\subsection{Parametric Neural Networks}
\label{subsec:pnn}
A \textit{parametric neural network} (pNN) \cite{baldi2016parameterized, cms2017search, cms2019qli} is a neural network architecture that leverages an additional input (in our case the \textit{mass} of the hypothetical particle) to replace many individual classifiers, and potentially even improve their classification performance. Let be $x$ the input features, $m$ the generated mass of the signal (or the signal mass hypotheses), and $\theta$ a set of learnable weights (or parameters). A pNN can be denoted as $f_\theta(x, m)$, i.e. as a learnable function of both the input features $x$ and the \textit{mass feature} $m$. A canonical neural network, instead, would be denoted as $f_\theta(x)$, depending only on the input features. The Baldi's pNN \cite{baldi2016parameterized} first concatenates $x$ with $m$, then applies five dense layers each with $500$ units and activated by ReLU, after that a final dense layer with sigmoid non-linearity outputs the predicted class label. Such architecture results in about $1$M learnable parameters.\\
\\
Indeed the idea of "parametric" is not new, as in other fields of machine learning, like \textit{imitation learning} \cite{codevilla2018end}, \textit{multi-task and meta-learning} \cite{finn2017model}, \textit{unsupervised reinforcement learning} \cite{eysenbach2018diversity}, and \textit{deep generative models} \cite{mirza2014conditional}, is commonly called "conditioning". Here the general idea is to condition the learning (i.e. output) of a neural network on some additional representation $z$, in order to let the network's output change as $z$ varies. The vector $z$ -- called the \textit{task representation} -- can take various forms: ranging from a one-hot encoding to a dense embedding, or be a single discrete or continuous variable as well. In our case, the mass feature (i.e. the task representation) is a single scalar $m$ belonging to a finite set $\mathcal M=\{m_1, m_2,\ldots, m_M\}$ of mass hypotheses about the signal process we're interested in.\\
\\
This idea, whether called conditioning or parametrization, is promising in HEP since it may enable to replace (potentially many) individual classifiers with a unique classifier trained on all mass hypotheses. Thus, leveraging the \textit{sharing of weights} for more efficient learning, and \textit{distributed representations} shared among mass hypotheses for improved predictions, which we also found to be beneficial in \textit{low-data regimes}: i.e. when some of the data corresponding to certain masses, is \textit{imbalanced} compared to the most representative masses\footnote{This is often the case in practice, since when doing MC simulations to reproduce the data, some kind of events (for certain masses) are more frequently generated (thus being less rare) than others, naturally resulting in an \textit{imbalanced dataset}, that is imbalanced not necessarily with respect the class labels (since the background class is independently produced) but with respect the \textit{mass feature}.}. The authors also claim that, since the additional input would define a \textit{smoothly varying learning task}, a pNN would be able to \textit{smoothly interpolate} between such learning tasks. Ideally, this means that a pNN would be able to generalize (correctly classify events) beyond the mass hypotheses it was trained on, thanks to the additional mass feature.

In this case, our supervised dataset $\mathcal D$ have the form $\{(x, m, y)_i\}_{i=1}^N$, in which we have input features $x$ (i.e. the variables associated to each event), their mass hypothesis $m$, and the target class labels $y$ we aim to predict. For each signal mass hypothesis, $m_i\in \mathcal M$, we can \textit{slice} our dataset such that only the features and targets corresponding to mass $m_i$ are retained, i.e. $\mathcal D_{m_i} = \{(x, y)_j\ : m_j = m_i, \forall j = 1,\ldots, N\}$, where $N$ is the total number of events\footnote{From a ML jargon perspective we can equivalently refer to events as (ex)\textit{samples} or \textit{datapoints}, of some dataset.}. In this way, we obtain $|\mathcal M|$ datasets for which each of them can be used to train an individual classifier (but also to evaluate our pNN at single mass points). The pNN can replace all of them, and if trained on a subset of the mass hypotheses, $\bar{\mathcal M}\subset{\mathcal M}$, it is, in principle, able to automatically account for the missing masses ($\tilde m_j \in \mathcal M \smallsetminus \bar{\mathcal M}$) thanks to the interpolation capability (discussed in section \ref{subsec:interpolation}), which should work on novel intermediate mass points as well.

\subsection{Conditioning Mechanisms}
\label{subsec:cond_mech}
\begin{figure}[h]
    \centering
    \begin{subfigure}{0.53\textwidth}
        \includegraphics[width=\textwidth]{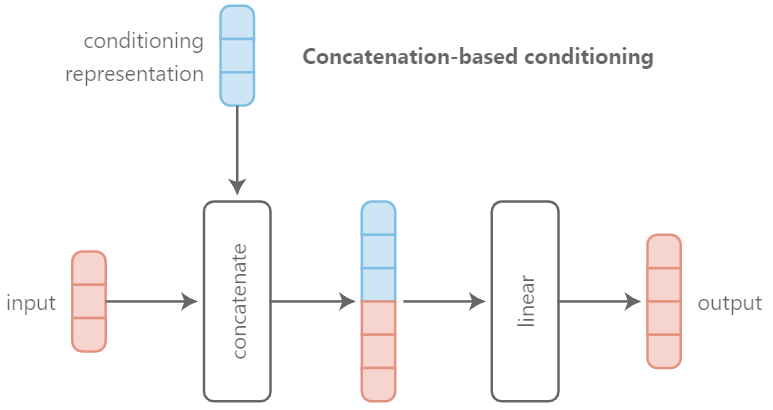}
        \caption{Concatenation-based conditioning}
    \end{subfigure}
    \begin{subfigure}{0.46\textwidth}
        \includegraphics[width=\textwidth]{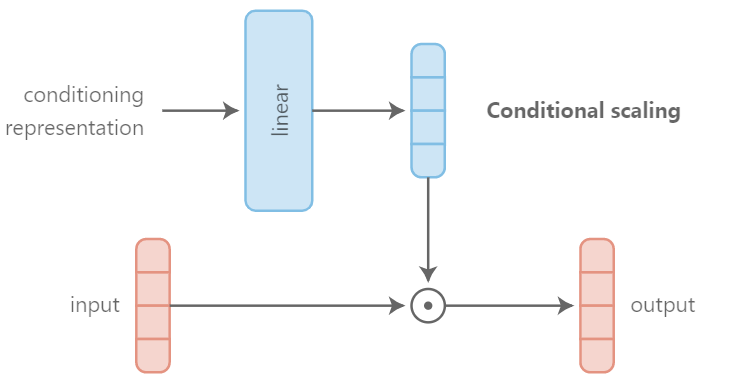}
        \vspace*{0.6mm}
        \caption{Conditional scaling}
    \end{subfigure}
    \caption{Two popular conditioning mechanism, that tuned out to be complementary. Reproduced from \cite{dumoulin2018feature-wise}.}
    \label{fig:conditioning}
\end{figure}

The authors of the original pNN \cite{baldi2016parameterized} utilize a simple conditioning mechanism: they just \textit{concatenate} the features with the mass (or task representation, in general) obtaining a new set of features, $\bar x = [x, m]$, going back to a standard feed-forward neural network formulation, i.e. $f_\theta(\bar x)$, that learns from an \textit{extended} set of features $\bar x$. Indeed, many arbitrarily-complex conditioning mechanisms exists, two of them (figure \ref{fig:conditioning}) are yet simple but powerful \cite{dumoulin2018feature-wise}:
\begin{itemize}
    \item \textbf{Concatenation-based conditioning:} the task or conditioning representation $m$ (our mass) is first concatenated along the last dimension (axis) of the input features $x$, and then the result is passed through a linear layer. Notice that in the original pNN, the linear layer after concatenation is missing.
    \item \textbf{Conditional scaling:} a linear layer first maps the conditioning representation $m$ to a scaling vector $s$, to which follows an element-wise multiplication (Hadamard product) with the input features $x$, i.e. $x\odot s$.
\end{itemize}

These two conditioning mechanisms are widely applicable, although it's not yet clear in which case one mechanism is preferable to the other(s). Anyway, these two mechanisms can be both combined into a third one\footnote{An affine transformation, i.e. $y=Ax + b$, does not involve concatenation directly: turns out that concatenation-based conditioning is equivalent to \textit{conditional biasing}, in which the task representation is first mapped to a bias vector that is then added to the input, element-wise, effectively replacing a concatenation operation. Further details in \cite{dumoulin2018feature-wise}.}: a \textit{conditional affine transformation} \cite{dumoulin2018feature-wise}, which motivates our new parametric architecture (refer to section \ref{subsec:affine_arch}).

\section{Datasets}
\label{sec:datasets}
In this section, we provide details about the two datasets we used to conduct our study.

\subsection{HEPMASS}
\label{subsec:hepmass}
\begin{figure}[t!]
    \centering
    \includegraphics[width=0.95\textwidth]{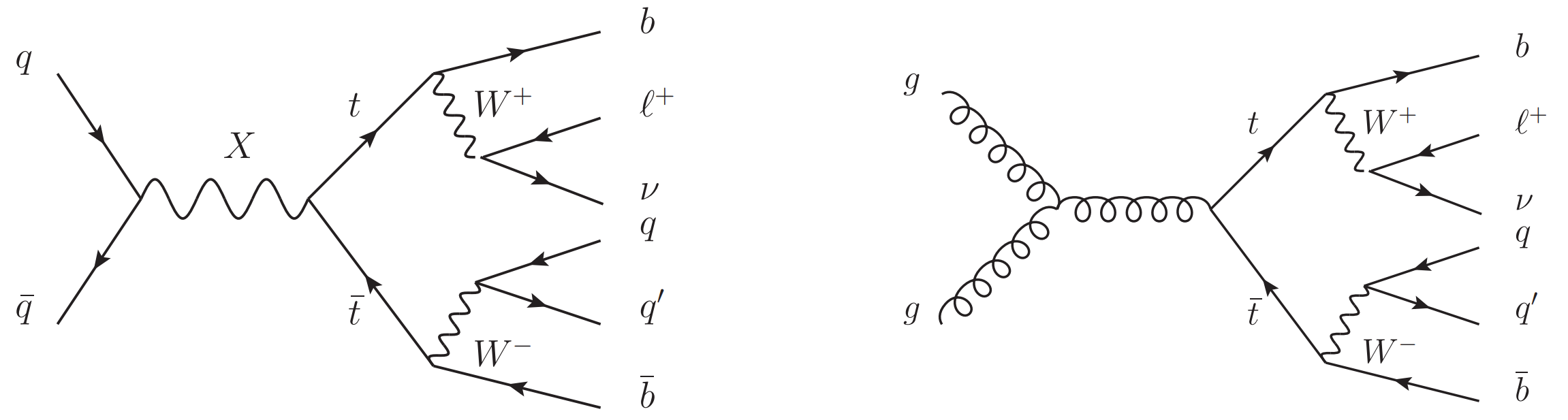}
    \caption{Feynman diagrams depicting the hypothetical particle decay: $X\to t\bar t\to W^+bW^-\bar{b}$. Reproduced from \cite{baldi2016parameterized}.}
    \label{fig:fey_diag}
\end{figure}

The \hepmass\ dataset \cite{baldi2015hepmass, baldi2016parameterized} was utilized by the pNN's authors to demonstrate their novel idea. The dataset contains $7$M training samples, and $3.5$M test samples.  The physical case under consideration is the search of a new particle $X$ with an unknown mass. This particle decays into a $t\bar{t}$ pair, and the final state consists in the most probable decay product: $t\bar{t}\rightarrow W^+bW^-\bar{b}\rightarrow qq'bl\nu\bar{b}$. The dominant background considered for this specific signal is the standard model $t\bar{t}$ production, identical in the final state but different in kinematics due to the absence of the $X$ resonance. The Feynman diagrams showing the signal and background processes are shown in Figure \ref{fig:fey_diag}. There are a total of 27 features (without considering the 28-th \textit{mass feature}) already normalized to have approximately zero-mean and unitary variance. Each datapoint $x^{(i)}\in\mathcal{D}$  can belong to either a signal process, with a mass hypotheses (in GeV) $m_X=\{500, 750, 1000, 1250, 1500\}$, or to a background process, where the mass feature is randomly sampled from $m_X$. Signal (background) samples are then labeled with class $1$ ($0$). Moreover, the two classes are perfectly balanced, and also each $\mathcal{D}_{m_i}$ is \textit{balanced}: containing the same amount of events for each $m_i\in m_X$. As discussed in the previous section, when training pNNs we want to pay attention to the balance of classes as well as the balanced of each $m_i$: this dataset avoids such issue. For further details about the data, refer to \cite{baldi2016parameterized}.

\begin{figure}[h]
    \centering
    \begin{subfigure}{0.45\textwidth}
        \includegraphics[width=\textwidth]{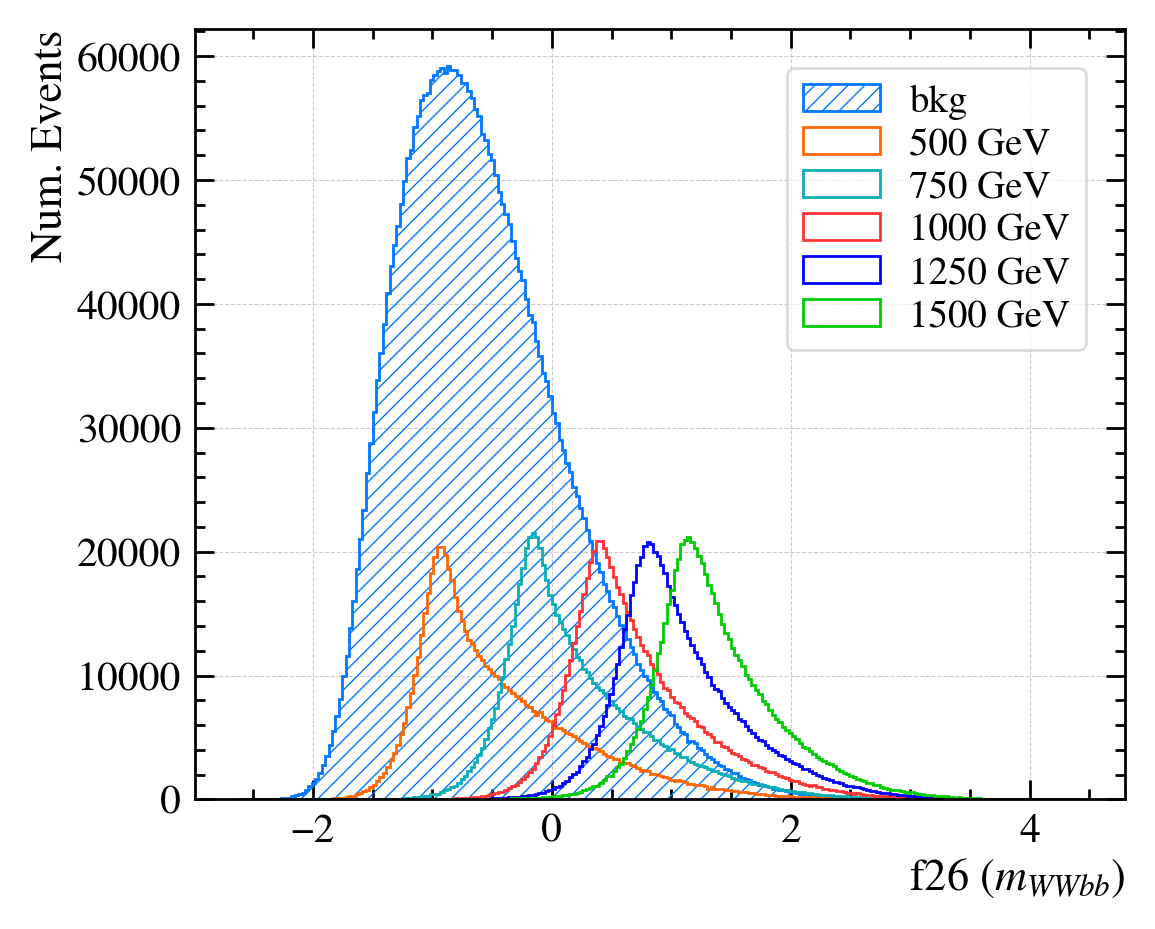}
        \caption{Distribution of the normalized feature \texttt{f26}: the whole background class is represented in blue, whereas each signal mass hypothesis with a different color.}
    \end{subfigure}
    \hfill
    \begin{subfigure}{0.51\textwidth}
        \includegraphics[width=\textwidth]{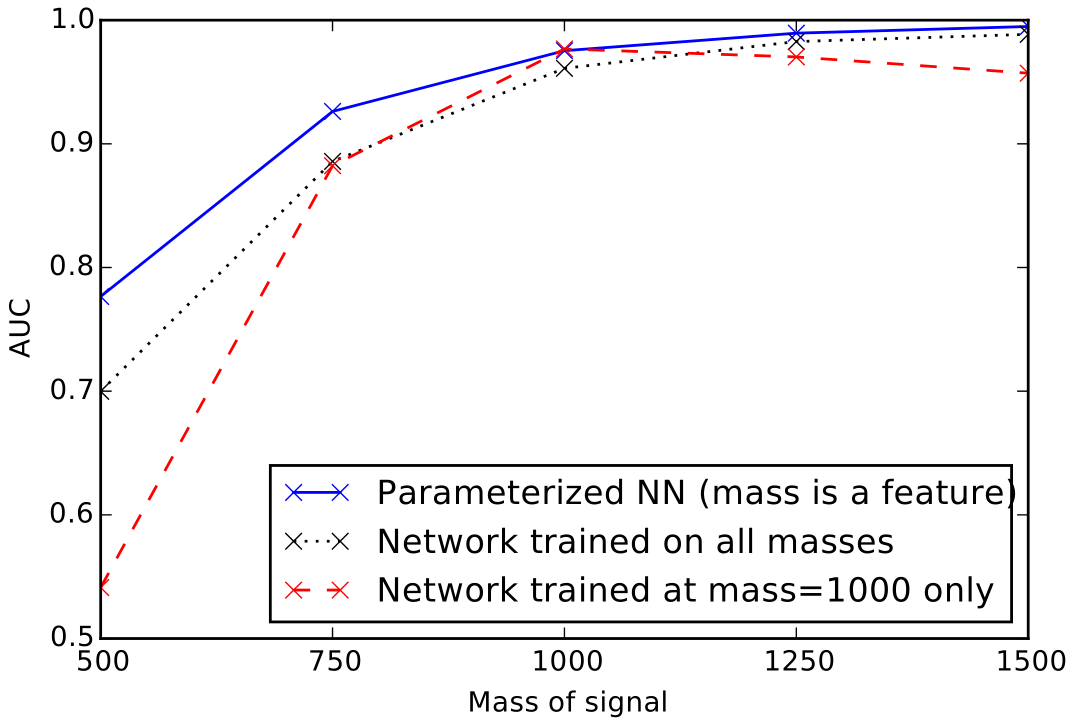}
        \caption{Per-mass AUC (area under the curve) of the ROC (receiver operating characteristic) curve: figure from \cite{baldi2016parameterized}.}
    \end{subfigure}
    \caption{The plot (a) shows the \textit{reconstructed mass} ($m_{WWbb}$) of the simulated decay ($X\to t\bar t\to W^+bW^-\bar{b}$), depicted in \hepmass. We observe that the background spans the entire mass range, mostly overlapping the signal at $m_X = 500$ and $750$ \gev. This fact also motivates why the authors' results (presented in plot b), and also our own, exhibit a neat loss in AUC below $750$ \gev, especially at $500$ \gev.}
    \label{fig:mass_shift}
\end{figure}

\begin{figure}[h]
    \centering
    \begin{subfigure}{0.24\textwidth}
        \includegraphics[width=\textwidth]{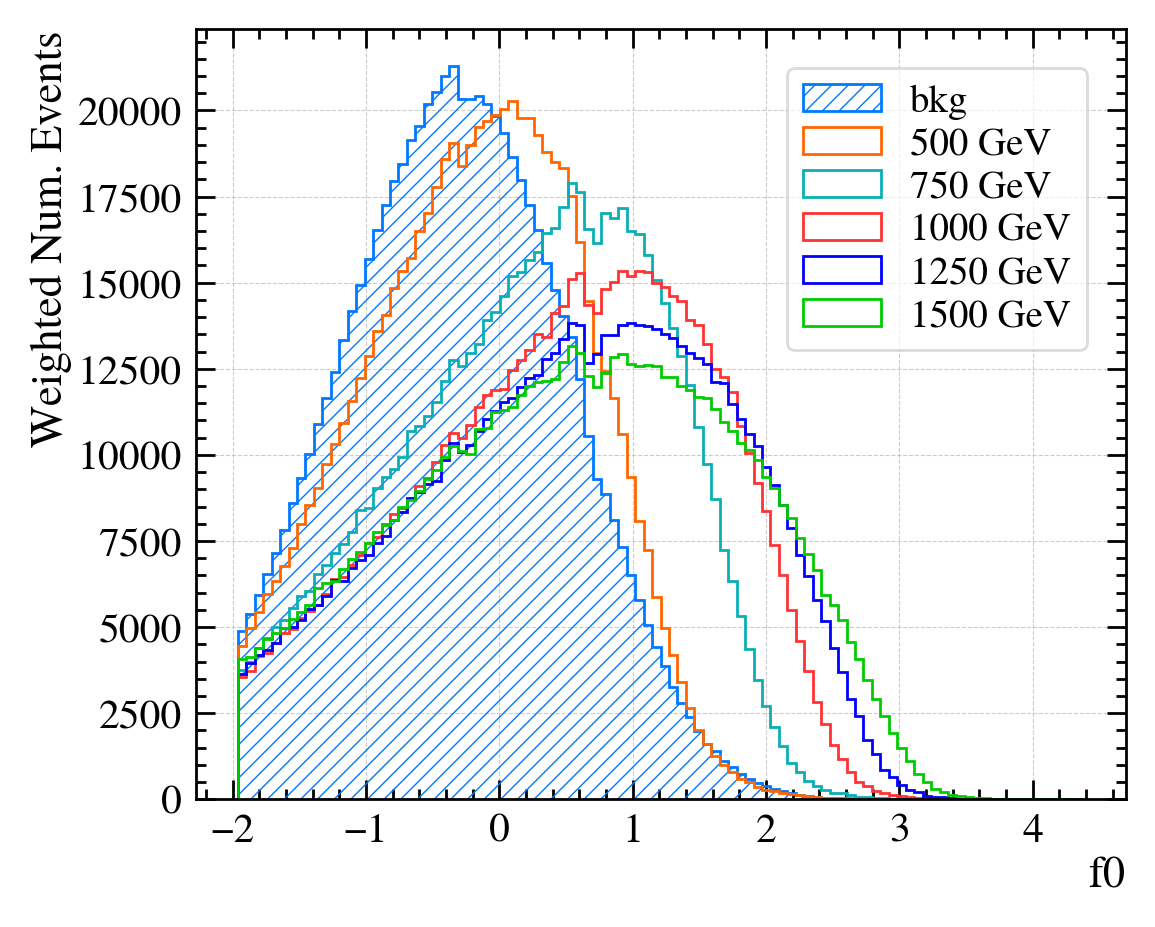}
        \caption{Feature \texttt{f0}}
    \end{subfigure}
    \begin{subfigure}{0.24\textwidth}
        \includegraphics[width=\textwidth]{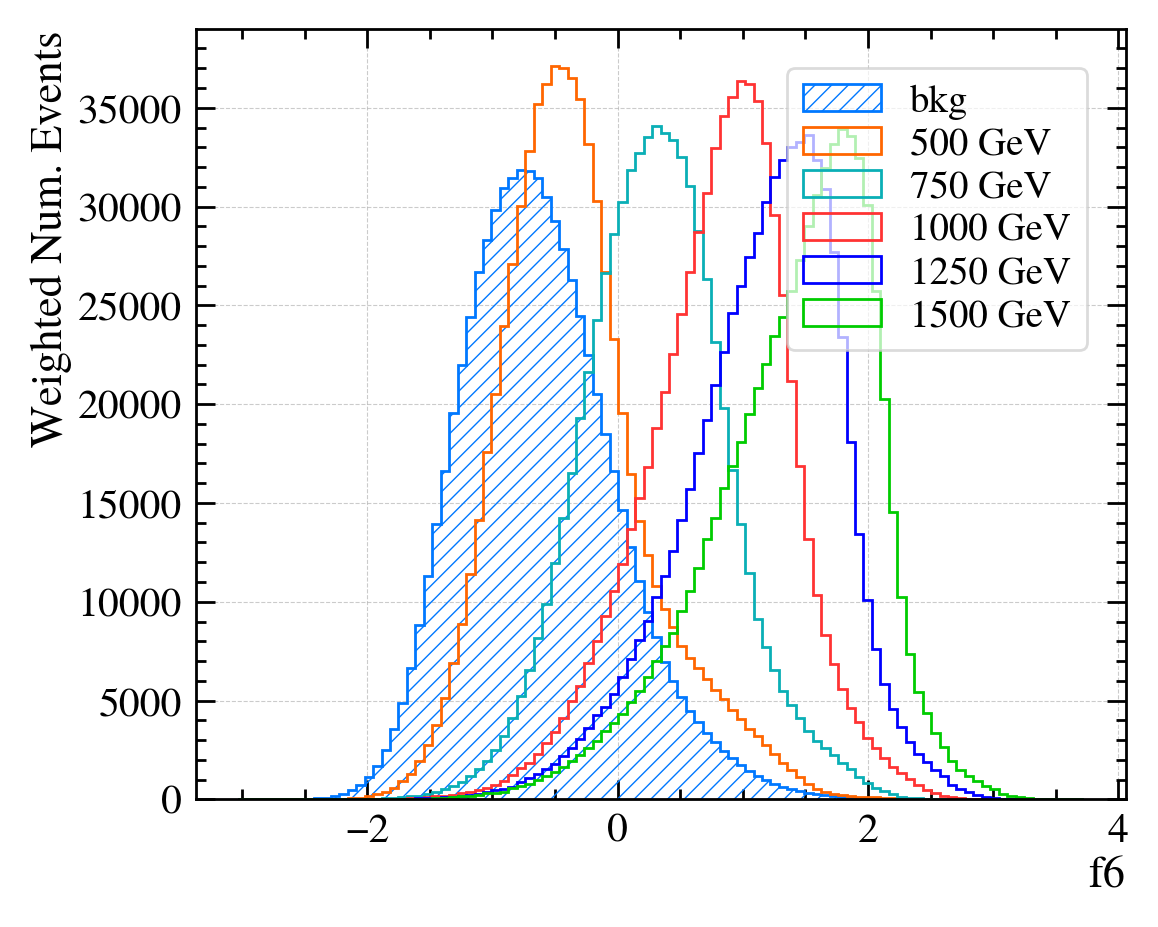}
        \caption{Feature \texttt{f6}}
    \end{subfigure}
    \begin{subfigure}{0.24\textwidth}
        \includegraphics[width=\textwidth]{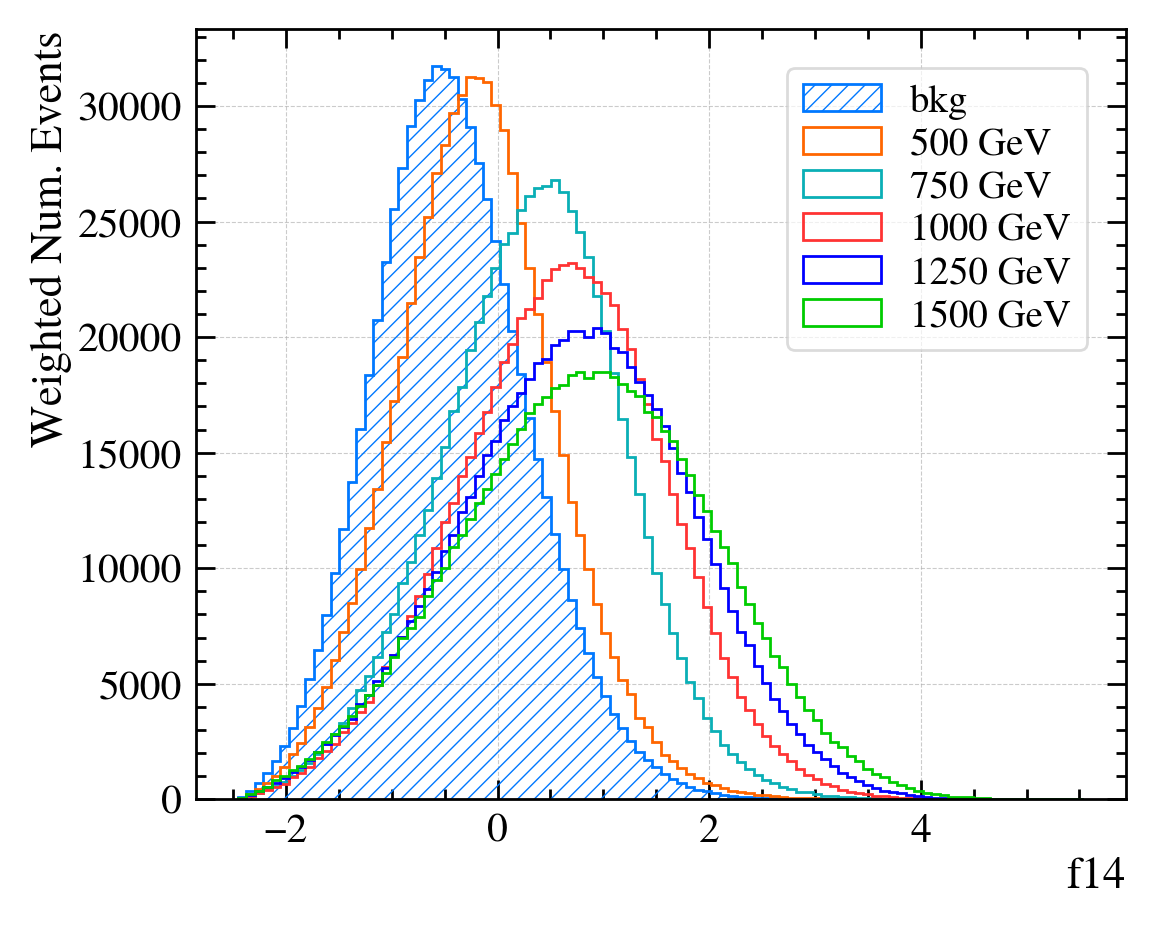}
        \caption{Feature \texttt{f14}}
    \end{subfigure}
    \begin{subfigure}{0.24\textwidth}
        \includegraphics[width=\textwidth]{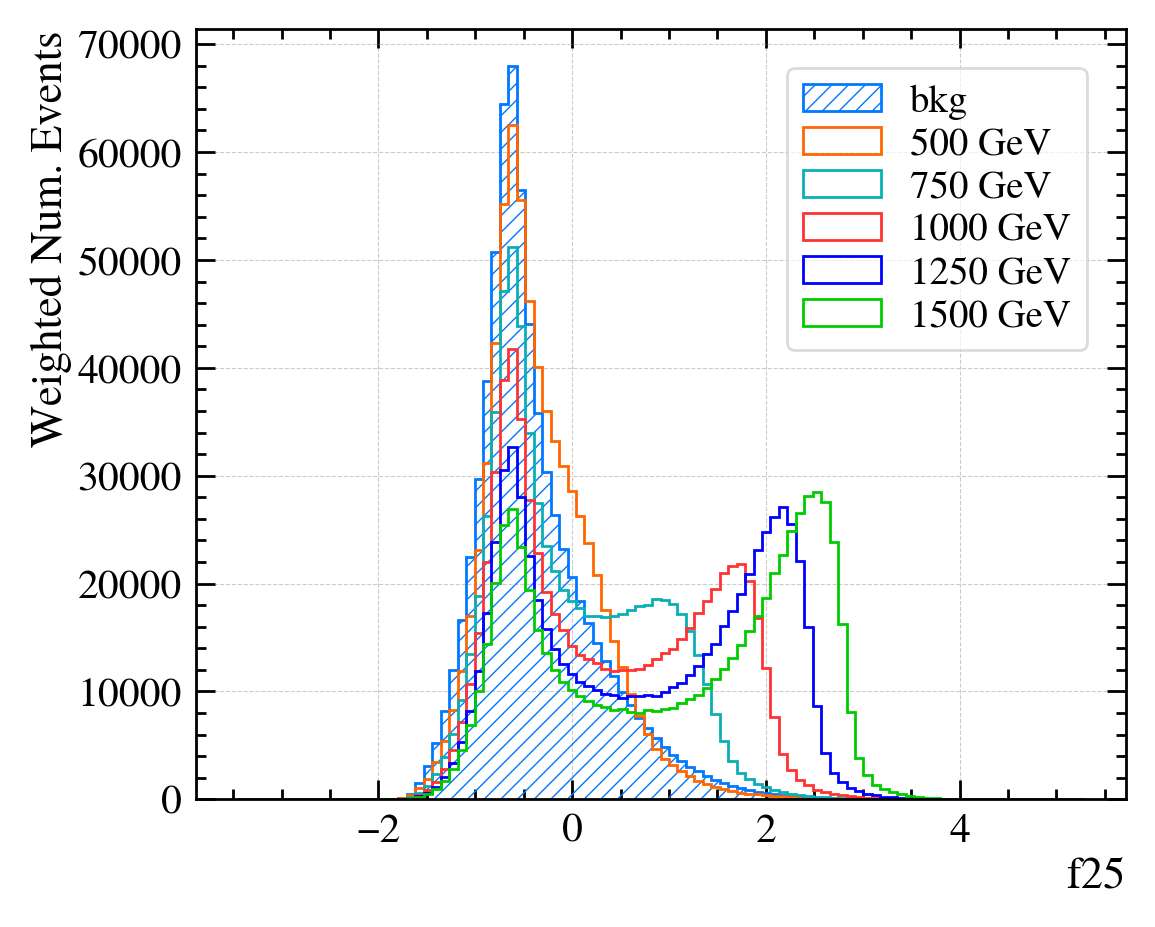}
        \caption{Feature \texttt{f25}}
    \end{subfigure}
    
    \caption{Some features in which we can clearly observe how easily the distribution of each mass hypothesis drifts away from the background. In particular, the distribution of $m_X = 500$ almost completely overlaps with the background, each time telling us that these events are the most difficult to classify (as shown in figure \ref{fig:mass_shift}b). The features shown here are all normalized.}
    \label{fig:mass_features}
\end{figure}

\begin{figure}[h]
    \centering
    \begin{subfigure}{0.30\textwidth}
        \includegraphics[width=\textwidth]{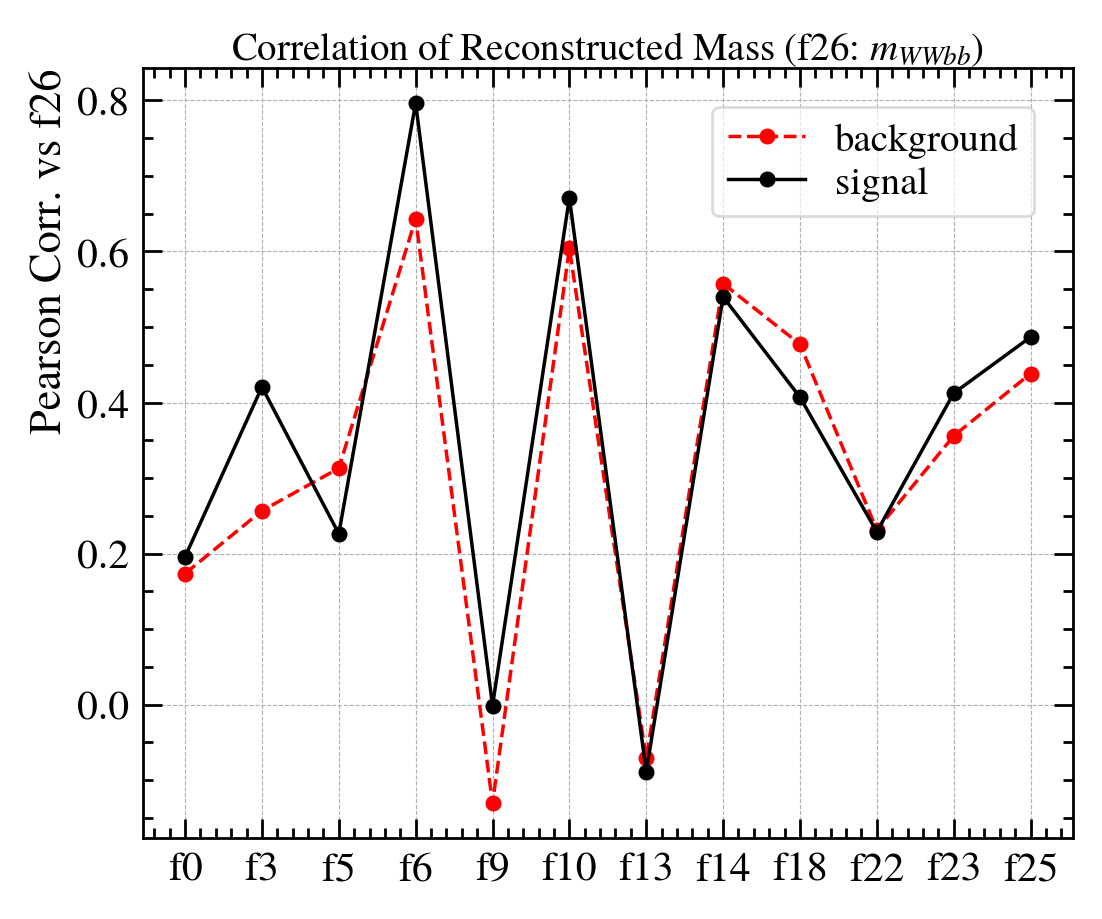}
        \caption{Correlation of both signal and background events with the reconstructed mass ($m_{WWbb}$) of the particle decay.}
    \end{subfigure}
    \hfill
    \begin{subfigure}{0.32\textwidth}
        \includegraphics[width=\textwidth]{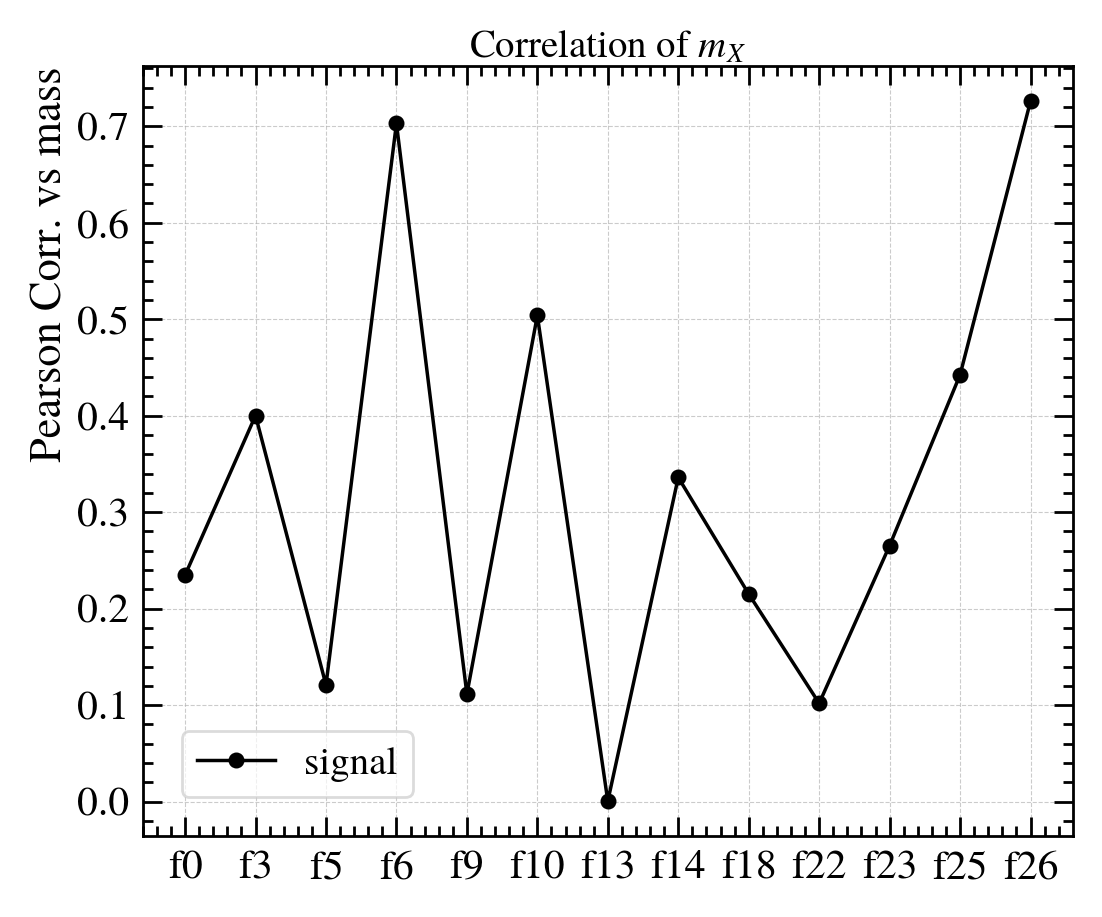}
        \caption{Correlation of the signal events with the mass hypotheses ($\mass$).}
    \end{subfigure}
    \hfill
    \begin{subfigure}{0.32\textwidth}
        \includegraphics[width=\textwidth]{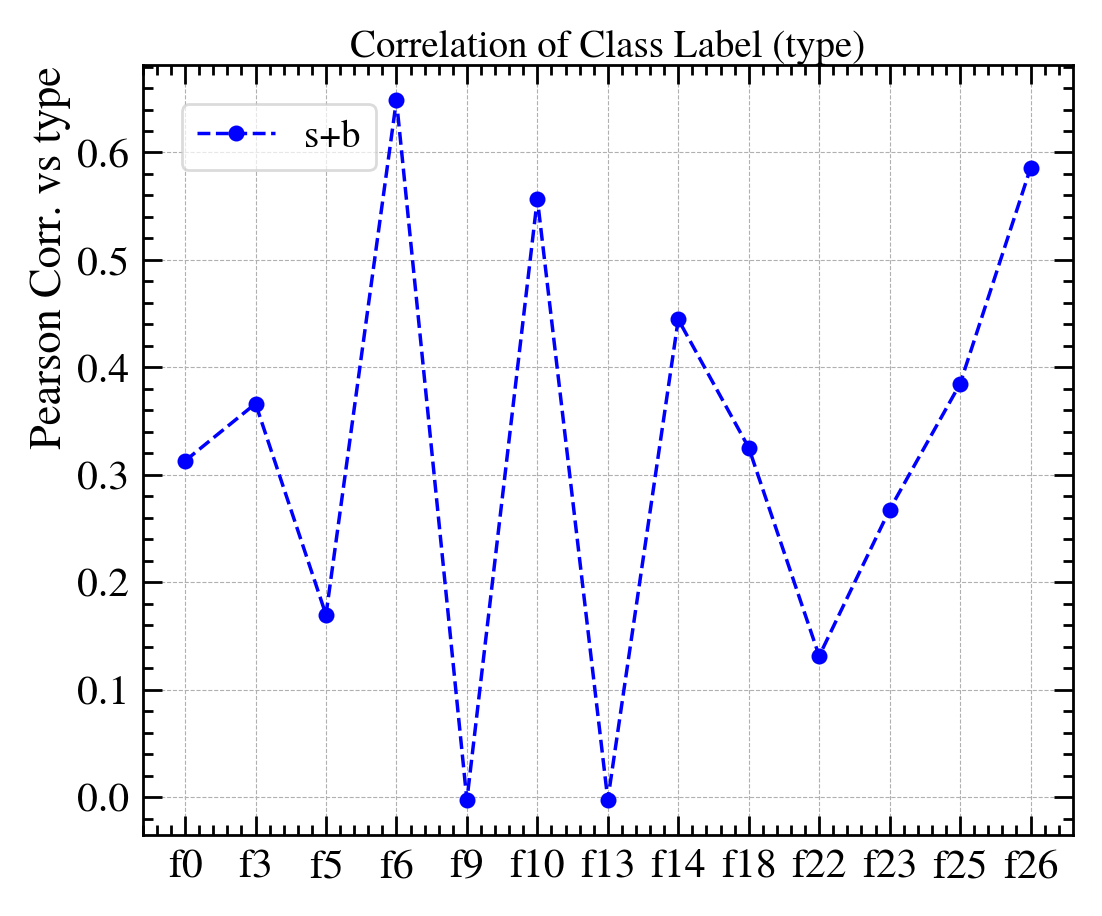}
        \caption{Correlation of all the events with the class-label (\texttt{type}).}
    \end{subfigure}
    \caption{Pearson correlations of the training variables of \hepmass\ (train-set).}
    \label{fig:hep_corr}
\end{figure}

By studying the distribution of each feature we can deduce three major things:
\begin{enumerate}
    \item The background is unique and covers the signal, although partially (figure \ref{fig:mass_shift}a).
    \item Consequently, the signal's events at $m_X=500,750$ GeV are the most difficult to separate out from the background, since the features distribution is mostly completely overlapped with the background's one. This explains why, in figure \ref{fig:mass_shift}b, the AUC is considerably lower at $500$ GeV, while being almost perfect for $1500$ GeV.
    \item By only considering some features (figure \ref{fig:mass_features}) a classifier (even simple) can easily tell which event belongs to the signal-class or not, thanks to these features being highly correlated with the class label: figure \ref{fig:hep_corr}.
\end{enumerate}

\subsection{HEPMASS-IMB}
\label{subsec:hep_imb}
Since the \hepmass\ dataset is rather simple, leaving almost no room for improvement, we decided to \textit{imbalance} the dataset by hand in order to being able to demonstrate novel methods for improving pNNs: we call this new dataset, derived from it, \texttt{HEPMASS-IMB} \cite{anzalone2022hepimb}. In particular the dataset is \textit{doubly-imbalanced}: there is \textit{class-imbalance} with respect to the class label, and \textit{mass-imbalance} with respect the theoretical parameter ($m_X$), as well. A comparison between the two dataset is depicted in table \ref{tab:dataset_comparison}.

\begin{figure}[h]
    \centering
    \begin{subfigure}{0.50\textwidth}
        \includegraphics[width=\textwidth]{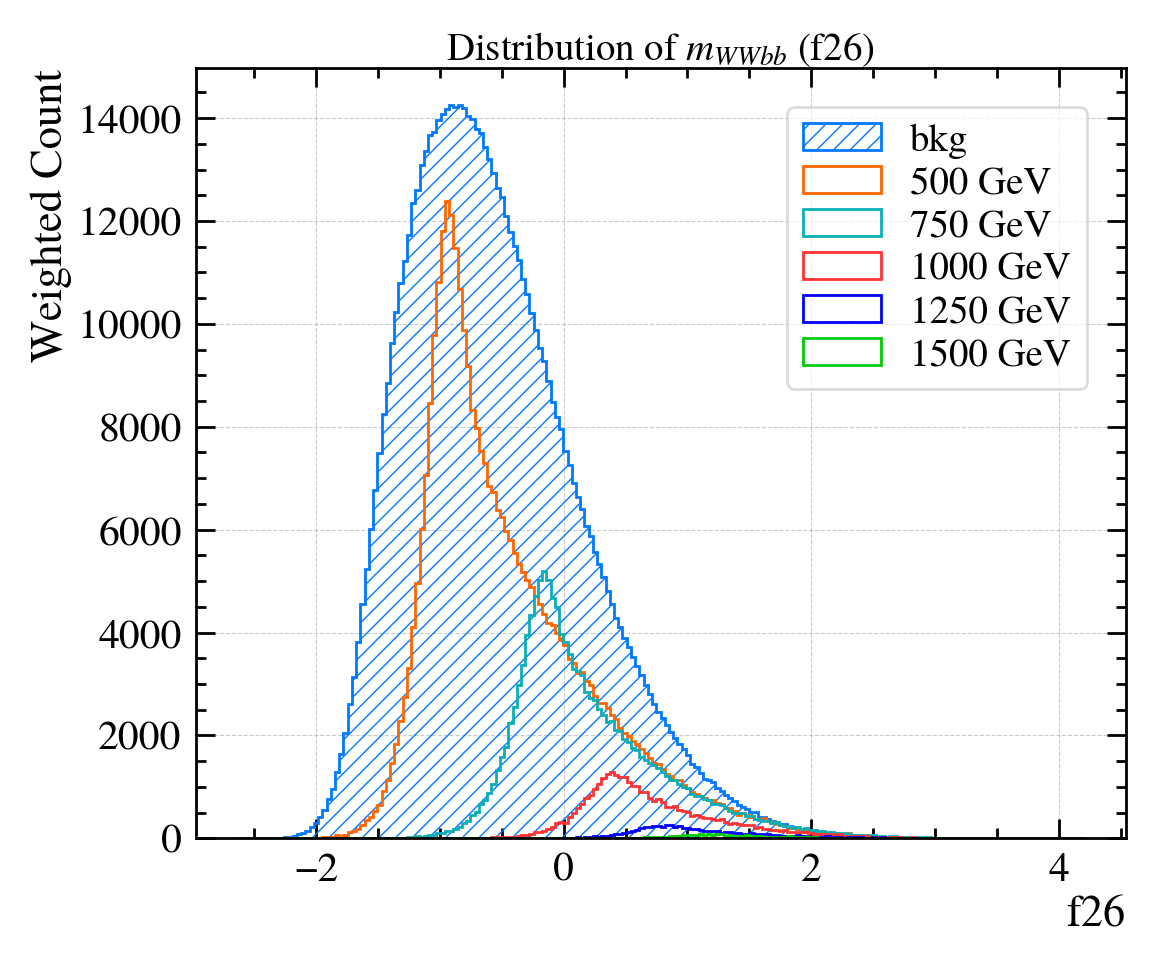}
        \caption{Resulting distribution of the reconstructed mass after having imbalanced \hepmass. Compared to figure \ref{fig:mass_shift}a, we can see how less the signal is: the background has been weighted by $1/5$, for visualization purpose.}
    \end{subfigure}
    \hfill
    \begin{subfigure}{0.47\textwidth}
        \includegraphics[width=\textwidth]{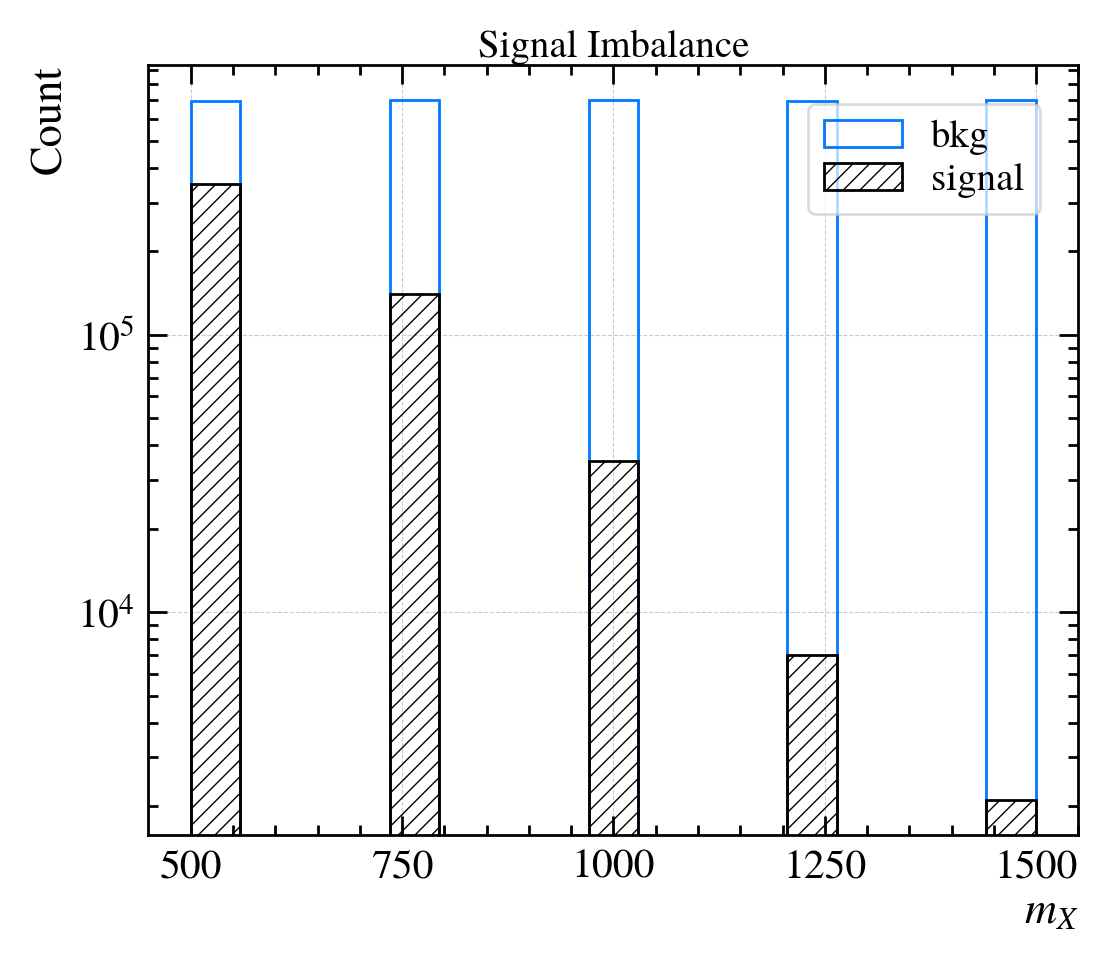}
        \caption{Imbalance of all the $m_i\in\mass$ signal mass hypotheses: log-scale. We can see that the imbalance ratio grows as $m_i$ increases, for a maximum difference of few orders of magnitude.}
    \end{subfigure}
    \caption{Overall class and mass imbalance of \hepimb.}
    \label{fig:hepimb_imbalance}
\end{figure}

The way we imbalance the dataset is as follows. We first take all the background events (without any change), and sub-sample (without replacement) only the signal, differently at each $m_i\in\mass$. In particular, we select: $350$k (for $m_X = 500$), $140$k (for $m_X = 750$), $35$k (for $m_X = 1000$), $7$k (for $m_X = 1250$), and lastly $2$k events for $m_X = 1500$; for a total of almost $534$k signal events. Indeed, we only imbalance the train-set of \hepmass, leaving its test-set as it was provided by the authors \cite{baldi2015hepmass}. In such way we are able to simulate a double imbalance of both class-labels and signal mass hypotheses (figure \ref{fig:hepimb_imbalance}), that resembles more the imbalance found in real-world dataset of Monte-Carlo simulated particle decays. 

\begingroup
    \setlength{\tabcolsep}{10pt}
    \renewcommand{\arraystretch}{1.5}
    
    \begin{table}[ht]
        \centering
        
        \begin{tabular}{ccc}
            \toprule
            \multirow{2}{*}{\textbf{Characteristic}} & \multicolumn{2}{c}{\textbf{Dataset}} \\ 
            & \multicolumn{1}{c}{\hepmass\ \cite{baldi2015hepmass}} & \multicolumn{1}{c}{\hepimb\ (our)} \\
            \midrule
            \# Samples & 7M (+ 3.5M test) & 4M (+ 3.5M test) \\
            Class imbalance (\%) & None & 13 (signal) / 87 (bkg) \\
            Signal imbalance & None & up to 166$\times$ \\
            Mass hypotheses &  5 (equally spaced) & // \\
            \# Variables & 27 & // \\
            Bkg distribution* & identical (fixed) & // \\
            Signal process & $X\to t\bar t\to W^+bW^-\bar b\to qq'b\ell\nu\bar b$ & // \\
            Bkg process & $t\bar t\to W^+bW^-\bar b\to qq'b\ell\nu\bar b$ & // \\
            \bottomrule
        \end{tabular}
        \caption{Datasets comparison. As we can see, our modification of \hepmass\ adds more difficulties to be addressed. This allowed us to discover many weaknesses of pNNs. (*) Design decisions about how to best distribute the mass of background events are depicted in section \ref{subsec:bkg}, going beyond what's fixed in the datasets.}
        \label{tab:dataset_comparison}
    \end{table}
\endgroup

\clearpage

\section{Method}
\label{sec:design_choices}

Having the extra \textit{mass feature} as input gives us an additional degree of freedom for the design of classifiers. In this section we study different design decisions about \textit{network architecture}, \textit{background distribution}, and \textit{training procedure}. 

\subsection{The Affine Architecture}
\label{subsec:affine_arch}
Baldi et al \cite{baldi2016parameterized} utilized a regular \textit{feed-forward} design for their parametric network, which we will refer to (vanilla) \textit{pNN}, by just concatenating the mass feature $m$ with the input features $x$ right after the input layer. Here we propose a novel conditioning (or parametrization) scheme to better exploit $m$, that is based on the two conditioning mechanisms described in section \ref{subsec:cond_mech}, namely: \textit{conditional scaling}, and \textit{conditional biasing} (equivalent to concatenation-based conditioning).

\begin{figure}[h]
    \centering
    \includegraphics[width=0.75\textwidth]{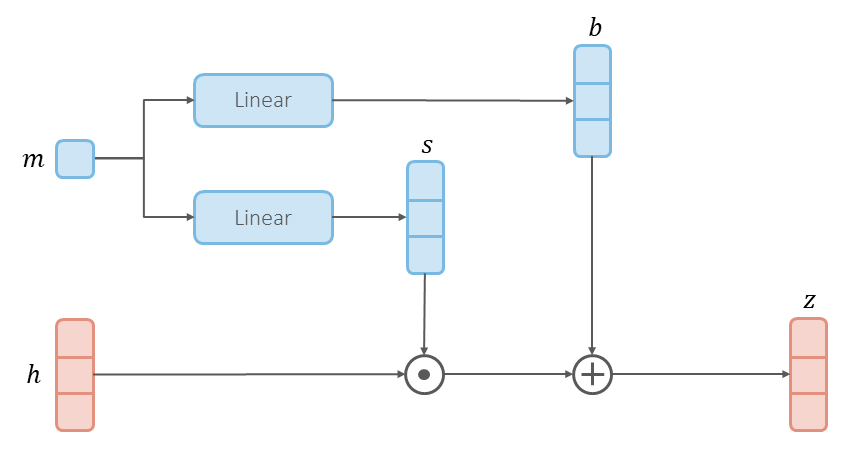}
    \caption{A schematic representation of the operations that build a single affine-conditioning layer. There are two inputs $h$ and $m$. The dimensionality of $m$ is expanded to match the dimensionality of $h$ through linear combinations, that yield scaling ($s$) and biasing ($b$) vectors that are, respectively, multiplied and added to $h$ in an element-wise fashion, resulting in the conditioned representation: $z$.}
    \label{fig:affine_cond_layer}
\end{figure}
\begin{figure}[h]
    \centering
    \includegraphics[width=\textwidth]{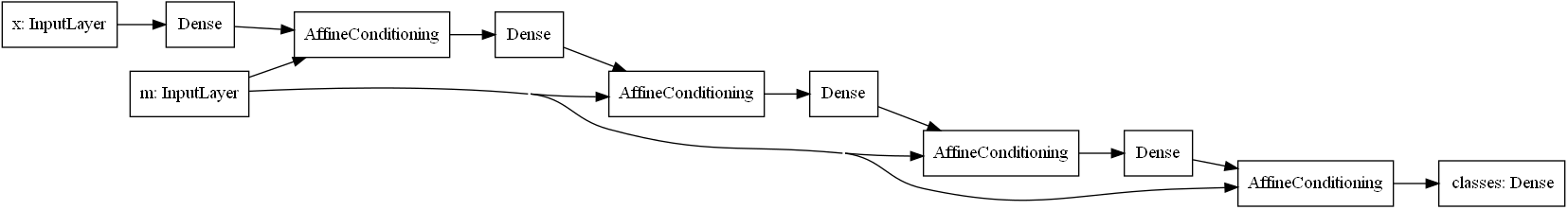}
    \caption{Diagram of the Affine architecture. In this picture the relation between the inputs and the layers can be better understood. At the beginning, the input features $x$ are only fed to the first \texttt{Dense} layer; the mass feature $m$, instead, is only provided to each \texttt{AffineConditioning} layer. \texttt{Dropout} layer are omitted for clarity.}
    \label{fig:affine_arch}
\end{figure}

We propose an \textit{Affine Parametric Neural Network} (AffinePNN) architecture, that relies on multiple \textit{affine-conditioning layers} (figure \ref{fig:affine_cond_layer}) instead of simply concatenating the mass at the beginning of the network. Such a layer takes two vectors $h$ and $m$ as input, where $h$ can be the features $x$ (if the layer is directly applied on the inputs) or the previous layer's output, and $m$ is the mass feature. Assuming them to have dimensionality $D_h$ and $D_m$ (that in our case is just one), respectively, the layer applies an element-wise affine transformation (scaling and bias addition) on $h$, such that the output $z$ is a function of $m$. 
Considering vectors at a generic index $i$ of the input batch, we have:

\begin{equation}
    z^{(i)}=h^{(i)} \odot s_\phi\big(m^{(i)}\big) + b_\psi\big(m^{(i)}\big),
    \label{eq:affine_pnn}
\end{equation}

where the dimensionality of $z^{(i)}$ is the same as $h^{(i)}$, i.e. $D_h$.
The scaling and biasing operations are defined as linear functions over the mass\footnote{In practice, these are implemented as two distinct \texttt{Dense} layers with linear activation.}: $s_\phi = W_\phi m^{(i)}$, and $b_\psi = W_\psi m^{(i)}$, where the learned weight matrices $W_\psi$ and $W_\phi$ have both shape $D_h\times D_m$, since the number of linear units have to match the dimensionality of $h^{(i)}$.
An AffinePNN interleaves such layers with ReLU-activated\footnote{$\mathrm{ReLU}(x) = \max(0, x)$, applied element-wise as usual.} dense layers: in figure \ref{fig:affine_arch} and table \ref{tab:affine_pnn}, an overview of the architecture is shown. In principle, the affine layers can be further generalized by introducing non-linear activation functions ($f$ and $g$) on the scaling $s_\phi$, and biasing $b_\psi$, such that: $z=h \odot f\big(s_\phi(m)\big) + g\big(b_\psi(m)\big)$.

\begingroup
    \setlength{\tabcolsep}{10pt}
    \renewcommand{\arraystretch}{1.5}
    
    \begin{table}[h!]
        \centering
        
        \begin{tabular}{ccc}
            \toprule
            \textbf{Layer} & \textbf{\# Units} \\
            \midrule
            Input ($x$) \\
            Dense + ReLU & 300 \\
            Input ($m$) \\            
            AffineConditioning & 300 \\
            Dropout & \\
            Dense + ReLU & 150 \\
            AffineConditioning & 150 \\
            Dropout & \\
            Dense + ReLU & 100 \\
            AffineConditioning & 100 \\
            Dropout & \\
            Dense + ReLU & 50 \\
            AffineConditioning & 50 \\
            Dropout & \\
            Dense + Sigmoid & 1 \\
            \bottomrule
        \end{tabular}
        \caption{Affine architecture. Each \texttt{AffineConditioning($m, z_i$)} layer takes two inputs: the masses $m$ input to the network, and the output $z_i$ of the previous \texttt{Dense} layer. For better performance \texttt{Dropout} layers are also included, but only after each \texttt{AffineConditioning} layer. Basically, the architecture is made of four \textit{blocks}, in which each block (\texttt{Dense} $\to$ \texttt{AffineConditioning} $\to$ \texttt{Dropout}) have, respectively, $300$, $150$, $100$, and $50$ hidden units, activated by ReLU non-linearity.}
        \label{tab:affine_pnn}
    \end{table}
\endgroup

In our preliminary experiments, we also evaluated some modifications of the network design shown in table \ref{tab:affine_pnn}. In particular, we tested various combinations of both activation function and weight initializer, finding that the ReLU activation paired with the default initialization scheme achieves the best performance: refer to section \ref{subsec:eval} for more details about the hyper-parameters. The last variation we tried, was the application of \textit{batch normalization} \cite{ioffe2015batch} after each affine-conditioning layer: the resulting trained network had similar classification performance, but at the cost of a slightly longer training (due to the overhead of the additional operations). One possible explanation is that since the network is not so deep (there are just four hidden dense layers), the gradient flow is not affected by either vanishing or exploding gradients, thus making batch normalization not fully necessary. This fact was confirmed by tracking the magnitude ($l_2$-norm) of the gradients during training. The same modifications were also tried on the pNN architecture, showing a similar behavior. Indeed, the choice of architecture-related hyper-parameters, like the activation function, weight initialization, number of layers (or blocks) and number of units, is usually dataset and problem dependent: what we found to be working here, is not said to be equally good in other circumstances. 

\subsection{Background's Mass Distribution}
\label{subsec:bkg}
In our reference work \cite{baldi2016parameterized}, the background's mass feature is \textit{identically distributed} as the signal's mass. In other analyses, background events receive a mass that is either: (1) \textit{randomly assigned} from the distribution of signal masses (only during training) \cite{atlas2021search}, or (2) \textit{uniform} in the interval of considered mass hypotheses. In general, we can analyze two situations: \textit{identical} and \textit{different} distribution of $m$ for background events only.
\begin{enumerate}
    \item \textbf{Identical distribution}: the mass feature for the $i$-th background event is assigned exactly to a value of the finite set $\mass = \{m_0, m_1, \ldots\, m_M\}$ of the signal mass hypotheses, selected randomly i.e. $m^{(i)}\sim\mass$. 
    \item \textbf{Different distribution}: the theoretical parameter ($m_X$) is used to define a probability distribution (e.g. \textit{uniform} in the interval $[\min(\mass), \max(\mass)]$, or \textit{Gaussian} being centered at each $m_i\in \mass$) that is \textit{diverse} from (going beyond) the finite set of values that $\mathcal{M}$ encodes. In this work, we only consider a uniform distribution $U(\mass_{\min}, \mass_{\max})$ for the mass feature, $m$. Therefore, the $i$-th background event will be assigned a mass feature by sampling randomly from such distribution, i.e. $m^{(i)}\sim U$.
\end{enumerate}

In both cases, we can \textit{fix} the values of the mass feature (only for background) in the dataset (e.g. by sampling $m$ only one time, and writing them to disk), or we can \textit{sample} them during training, repeatedly and differently at each mini-batch. So, we have two degrees of freedom (distribution type, and assignment strategy) that lead us to a total of four unique combinations: (1) \textit{identical fixed}, (2) \textit{identical sampled}, (3) \textit{uniform fixed}, and lastly (4) \textit{uniform sampled}. In our experiments we noticed that, without proper regularization, having a uniform mass feature for the background allowed the network to almost perfectly fit both the training and validation sets: this may be due to the introduction of an artificial correlation between the mass feature and the class label, that was exploited during training. Nevertheless, by regularizing the model enough generalization is still ensured.

We may further discuss an additional assignment strategy where the mass feature $m$ for the background can be also determined by means of \textit{mass intervals} (or bins), based on the underlying \textit{reconstructed mass} of the selected decay products. For example, if we consider $(150, 250)$ to be a specific mass interval centered around $200$ GeV, we can assign $m = 200$ as mass feature for each background event $x$ whose reconstructed mass is within the mass interval. In this work, such third assignment strategy have not been taken into account.

\subsection{Training Procedure}
\label{subsec:balanced_training}
Beyond the architecture and regularization of the parametric network, as well as the distribution of the mass feature $m$, we can make some further considerations about how to properly train such kind of neural networks in light of what we already know about the structure of our own data. In general, we known that the signal is arranged in $|\mass|$ groups (one for each $m_i$), and that the background is (eventually) composed of different processes. Therefore, beyond class labels, our data is naturally divided in \textit{sub-classes}: in terms of the signal generating mass (i.e. the various $m_i\in\mass$), and background processes\footnote{We can think of sub-classes as additional labels in our data.}. We can exploit such domain-knowledge to design a training procedure that embeds such \textit{inductive biases}.\\
\\
In particular, we can further notice that each sub-class may have its own unique \textit{frequency}, in terms of how much data samples fall into each sub-class, e.g. due to data imbalance: as shown in figures \ref{fig:mass_shift}a and \ref{fig:hepimb_imbalance}b. Such frequencies may bias the (parametric) network towards certain sub-classes, resulting in an overall sub-optimal fitting of the data.

We propose to mitigate this simply by \textit{balancing} each sub-class in a way that an equal number of events belongs to each of them. We call such approach \textit{balanced training}, that, without discarding or generating new data, can be easily implemented by balancing each \textit{mini-batch} during training, e.g. by sampling each sub-class in equal proportion. Specifically, we have:
\begin{itemize}
    \item \textbf{No balance:} The usual training procedure, in which the network is trained by experiencing the data as it is. This will be our baseline for comparison.
    \item \textbf{Class-only balance:} Only the class labels are balanced within each mini-batch. Considering two classes, we can balance them in two ways: (1) by associating \textit{sample or class weights} to each event, resulting in a weighted loss function, or (2) by sampling the events for a mini-batch such that half the batch is populated with samples belonging to the positive class (i.e. signal), and the other half with background samples (the negative class). We implement class-balancing by following the second option, as if the class weights were \textit{implicitly} provided to weight the loss function.
    \item \textbf{Signal balance:} Since the entire signal is generated at different values of $m_X$, we can build mini-batches such that there is an equal number of events for each $m_i\in\mass$. In practice, we take half the size $B$ of a batch, i.e. $B_s = B/2$, and then split that in even parts such that each $m_X = m_i$ is represented by exactly $B_s / |\mass|$ events. This implies that, at each mini-batch, all the signal mass hypotheses are always represented: 
    this may not occur, especially if the batch size is small, for the case of no class and background balancing.
    \item \textbf{Background balance:} Similarly to signal balancing, mini-batches are divided into equally-sized parts such that each part contains events that belong to a certain background process. Of course, such balancing strategy is only meaningful if our background comprises more than one process: not the case of \hepimb. Also in this case, each mini-batch will contain samples coming from all background processes.
    \item \textbf{Full balance
    \footnote{In our case, with only one background process, the signal-only and fully balanced training (in which half the batch size is reserved for signal, and the other half for background events) options are equivalent; not true, in general.}
    :} A combination of class, signal, and background balance. A batch is divided into two halves (each of size $B/2$) to ensure class balance. Then, one half will be signal balanced, and the other one is balanced according to the background. In this way, the network will always experience mini-batches that comprise all signal hypotheses, as well as all background processes.
\end{itemize}

\paragraph{Model Selection.} For every balancing procedure (even when the training is not balanced) and regardless the background's mass feature distribution, we always perform  \textit{validation} in the same way on a $25\%$ split of the \textit{original} (not batch-balanced) data. The AUC of the ROC is used as validation metric. The way we perform validation resembles the way the model is evaluated on the respective test-set (see section \ref{sec:evaluation}). In particular, for each $m_i\in\mass$ we take the corresponding signal samples $s$, and all the background $b$; for the latter only, we sample their mass feature $m$ from $\mass$: as in the identical (sampled) option, described in the previous section. Thus, the mass feature $m$, will be $m^{(s)} = m_i$ for the signal events, and $m^{(b)}\sim\mass$ for the background events. This is important to do, since: (1) balancing during validation will alter the value of the validation metric(s), as the original distribution of the data will be changed, and (2) regarding the mass feature distribution for the background, validating in a different way would lead to sub-optimal generalization performance of the selected model.  

\subsection{Preprocessing and Regularization}

In general, for our models we found out regularization to be beneficial for improved performance, and also (as we will see later) crucial for good interpolation. We utilize two well known regularization techniques: \textit{dropout} \cite{srivastava2014dropout}, and $l2$-\textit{weight decay}. For the affine architecture (section \ref{subsec:affine_arch}) we insert a \texttt{Dropout} layer after each affine-conditioning layer, thus zeroing random elements of the conditioned internal representation $z$: refer to Eq. \ref{eq:affine_pnn}. Instead, for the standard pNN architecture the \texttt{Dropout} layer is inserted after each ReLU activation. In both cases, we use a drop probability of $25\%$.
Lastly we apply $l2$-regularization on all learnable parameters of the network, but with different coefficients for weights and biases respectively.

Classification performances can be usually further boosted by properly normalizing the data input to the network. Recall that our data have the form $(x, m, y)$, in which: $x$ is a multi-dimensional vector of features, $m$ (the mass feature) corresponds to a 1-dimensional vector of values that make the network be "parametric", lastly $y$ is a vector of class-labels (either $1$ for signal or $0$ for background, in our case). Discarding $y$, we can normalize (or preprocess, in general) both $x$ and $m$ in the same way, as done by \cite{baldi2016parameterized} by means of \textit{min-max normalization}, or differently. In our case, as \hepmass\ is already standardized, we only normalize $m$ by just \textit{dividing} it by $1000$.

\section{Properties of PNNs}
\label{sec:properties}
We believe that parametric networks have many interesting properties beyond interpolation. In this section we attempt a first characterization of them, trying to better understand such kind of models.

\subsection{Interpolation}
\label{subsec:interpolation}
The interpolation capability of a parametric neural network is its ability to generalize towards novel mass points that lie between two known mass hypotheses, resulting in effective interpolation of events between them: we also use the term \textit{extrapolation} when the missing or novel mass points lie at the two extremes of the mass range, or even beyond them. In this case the generalization capability should be \textit{twofold}: the network is requested (1) to perform well on novel samples belonging to the known masses, and also (2) to correctly classify new events that belong to the missing hypotheses.
This means that a pNN capable of good interpolation should provide more accurate outcomes compared to the ones that would be obtained by interpolating the results of $M$ individual classifiers, instead.
We want our pNN to perform well even if some hypotheses are missing. So, how to be sure and ensure that our model has acquired such capability?

\paragraph{Factors.} We investigated several potential factors, including \textit{per-mass features distribution}, \textit{mass imbalance}, \textit{background distribution}, \textit{network regularization}, \textit{batch size}, and \textit{training procedure}, that may affect the interpolation capability of pNNs. Here we describe the most impactful:
\begin{itemize}
    \item \textbf{Per-mass features distribution}: Recalling from figure \ref{fig:mass_features}, a shift in the feature distribution has been observed. This tells us that some mass points are more difficult to classify than others. In fact such behavior is directly reflected on \textit{single-mass interpolation} and \textit{extrapolation} (i.e. when we train our model on just one mass less). In figure \ref{fig:single_mass_interp} we can observe how extrapolating $\mathcal{D}_{500}$ (plot d) is way more difficult than the other masses. This also suggests us that evaluating our model on only one mass less does not necessarily imply that our network will correctly interpolate or extrapolate everywhere, in general.
    
    \begin{figure}[h!]
        \centering
        
        \begin{subfigure}{0.32\textwidth}
            \includegraphics[width=\textwidth]{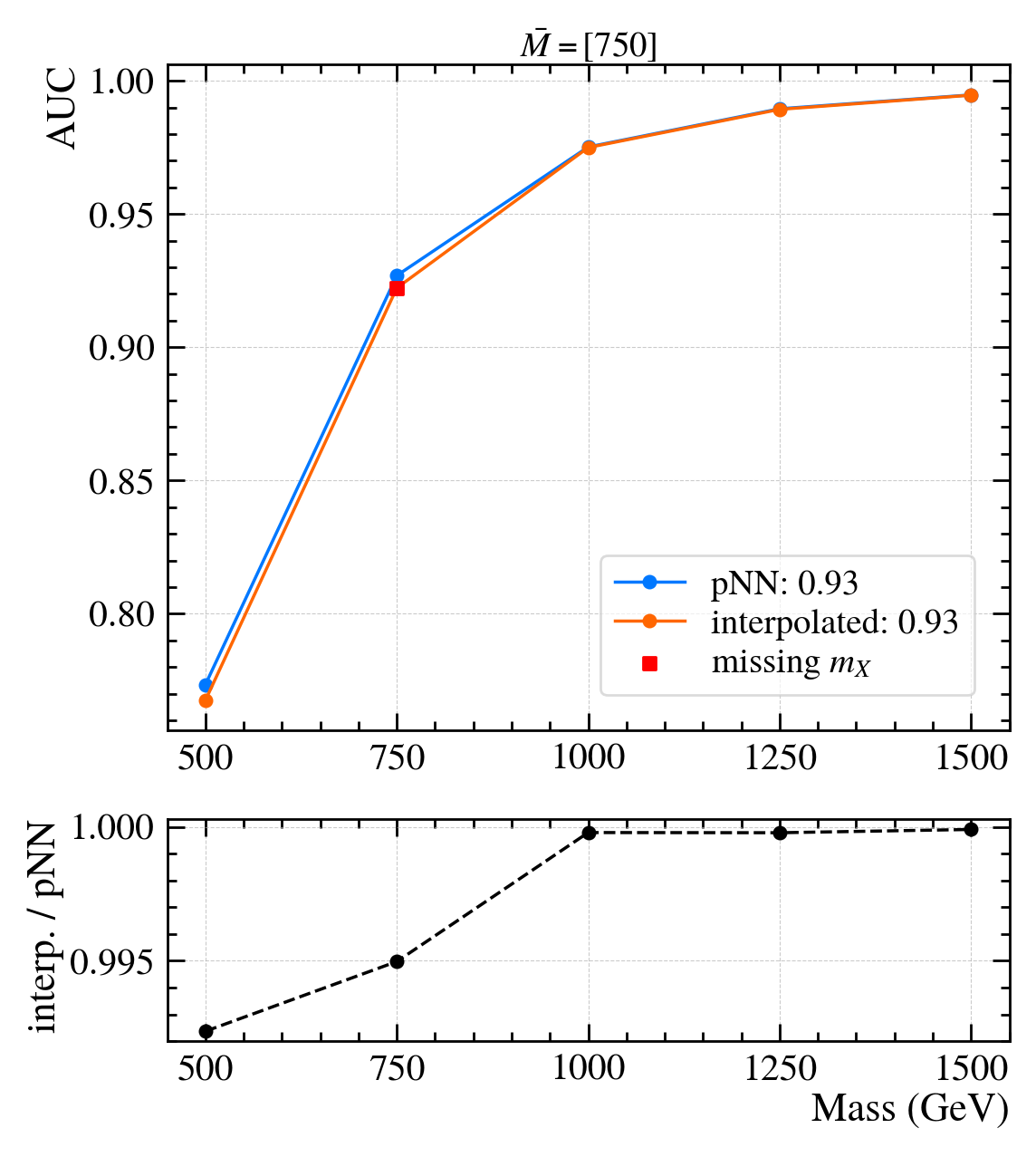}
            \caption{Interpolates $\mathcal{D}_{750}$}
        \end{subfigure}
        \hfill
        \begin{subfigure}{0.32\textwidth}
            \includegraphics[width=\textwidth]{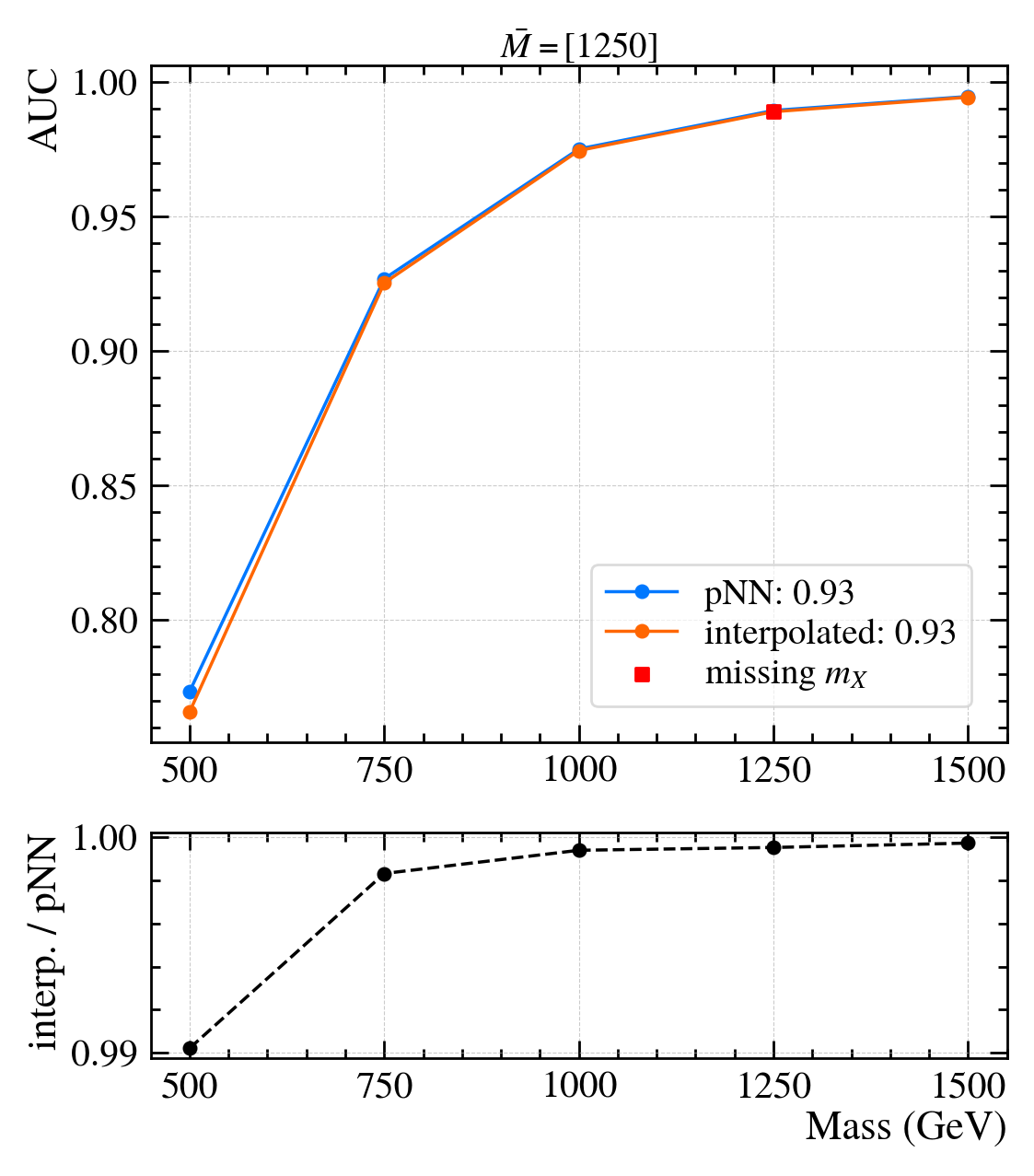}
            \caption{Interpolates $\mathcal{D}_{1250}$}
        \end{subfigure}
        \hfill
        \begin{subfigure}{0.32\textwidth}
            \includegraphics[width=\textwidth]{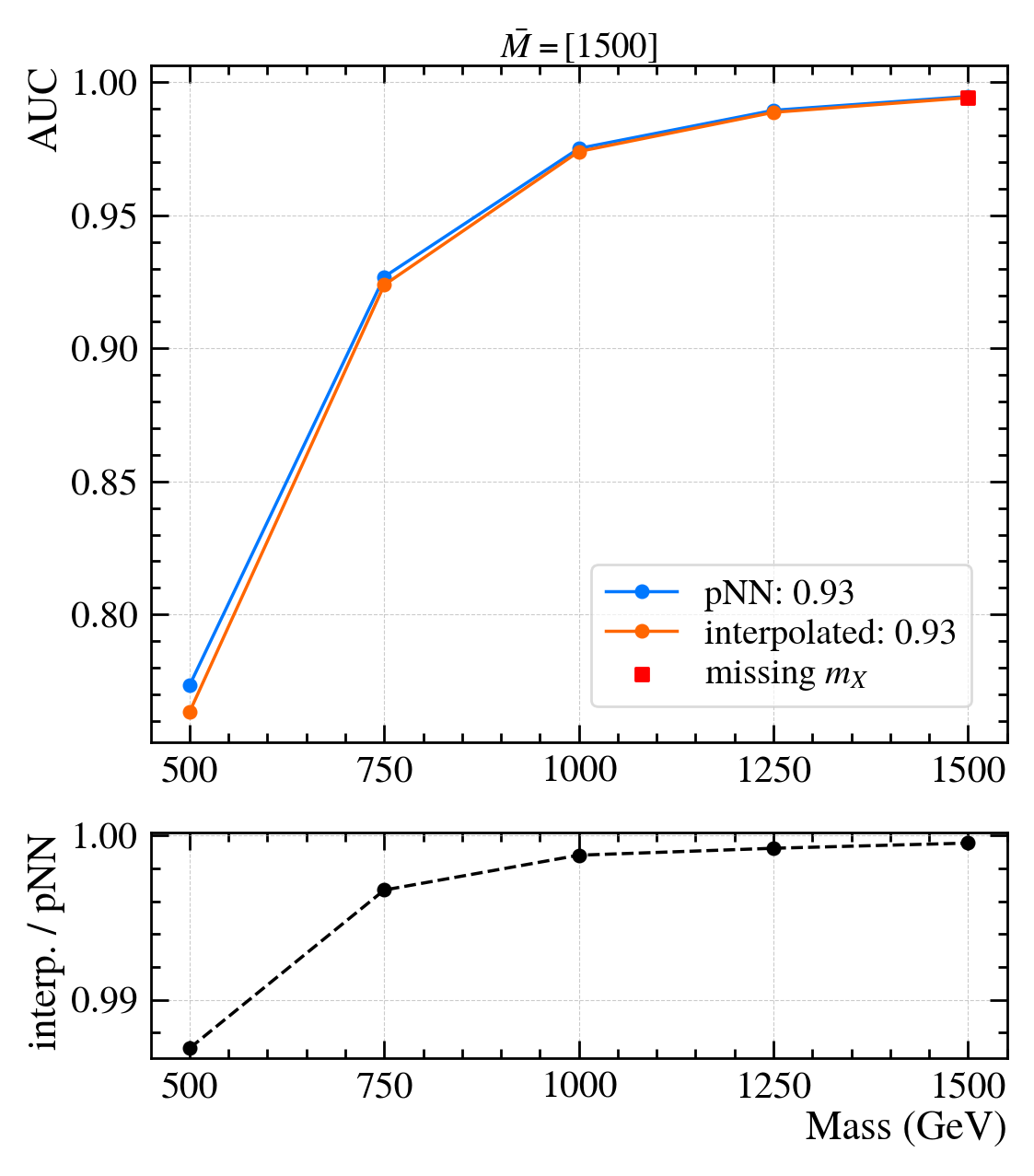}
            \caption{Extrapolates $\mathcal{D}_{1500}$}
        \end{subfigure}
        
        \begin{subfigure}{0.49\textwidth}
            \includegraphics[width=\textwidth]{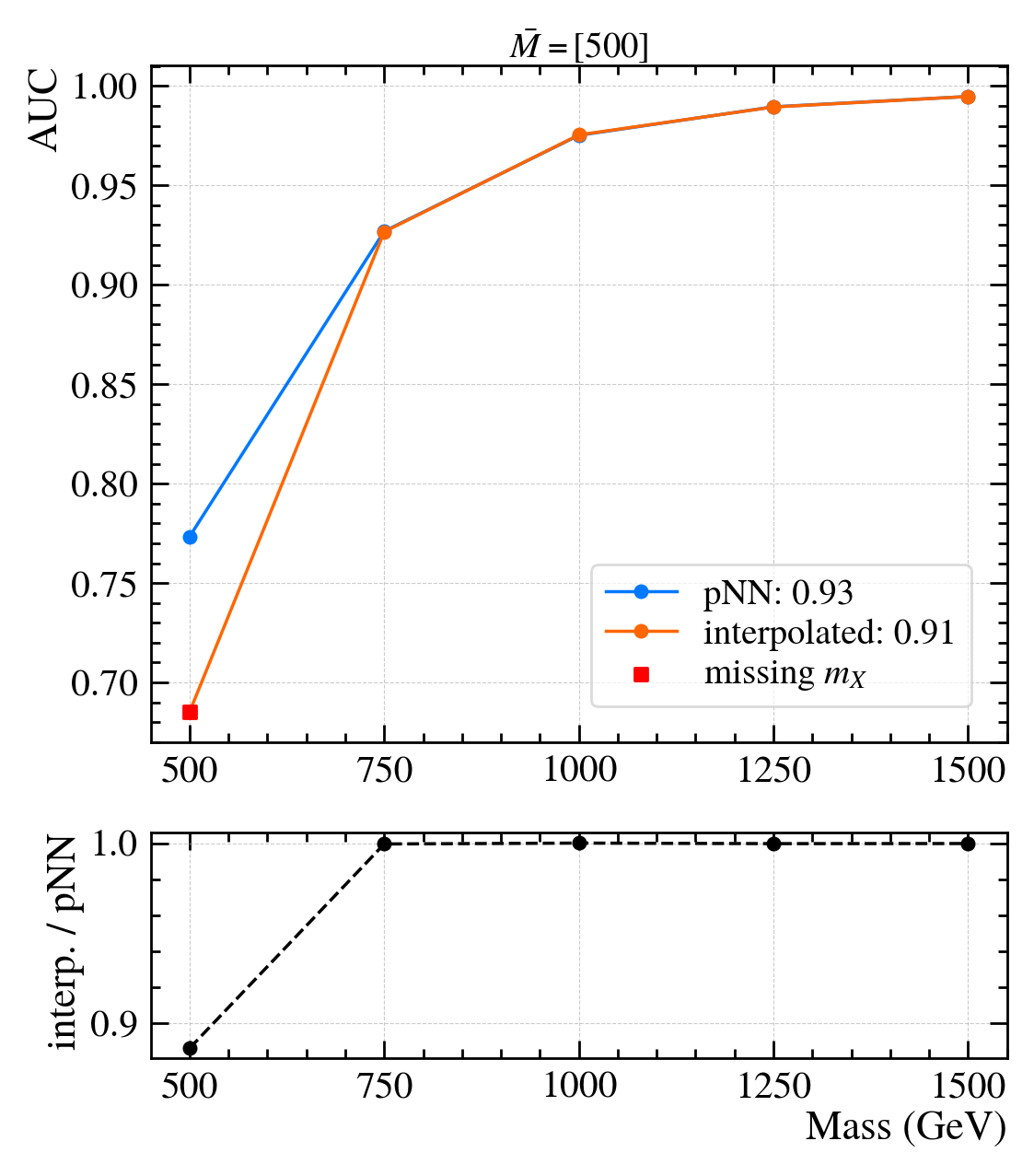}
            \caption{Extrapolates $\mathcal{D}_{500}$}
        \end{subfigure}
        \hfill
        \begin{subfigure}{0.49\textwidth}
            \includegraphics[width=\textwidth]{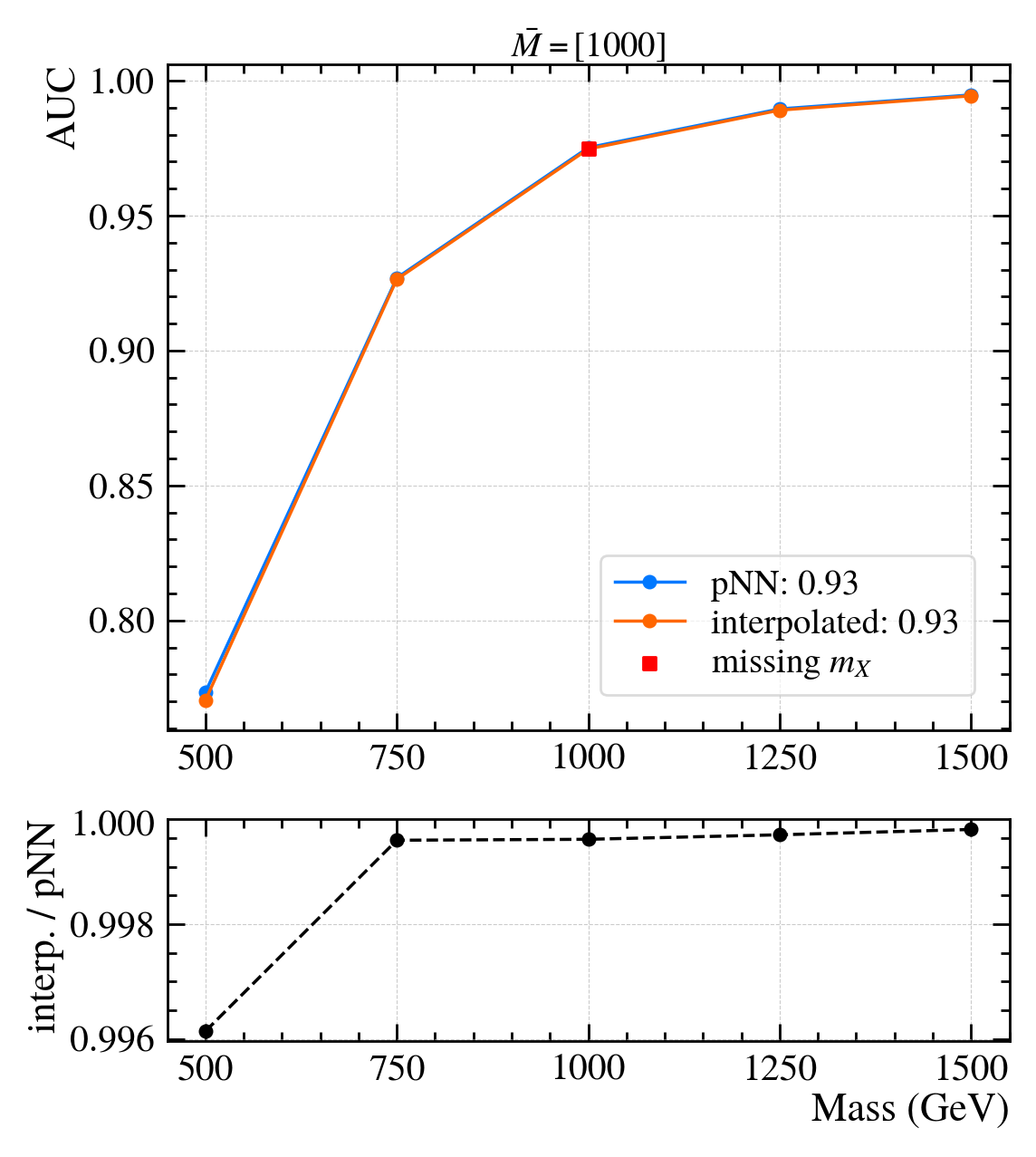}
            \caption{Interpolates $\mathcal{D}_{1000}$}
        \end{subfigure}
        
        \caption{Single-mass interpolation and extrapolation on \hepmass. The blue curve (same in all plots) represents the ROC's AUC of a pNN trained on all hypotheses. The orange curve depicts AUC performance for a pNN trained on a subset of the masses. The missing signal mass hypotheses are denoted by a red square. Lastly, below each plot is depicted the ratio between the orange and blue curves.}
    \label{fig:single_mass_interp}
    \end{figure}
    Another way to understand how much the similarity among masses helps (or avoids) our network at interpolating or extrapolating them, is to stress our model at \textit{extrapolating}: we train a pNN on just \textit{one} mass hypothesis, requesting it to extrapolate all the remaining ones. In figures \ref{fig:extrapolation}a and \ref{fig:extrapolation}b, we observe how easier is to predict the missing masses, being the model only trained on $m_X = 750$ or $m_X = 1000$. This fact seems to be (at least, partially) independent from the AUC achieved on such mass points: although on $\mathcal{D}_{1500}$ the highest AUC is obtained (plot \ref{fig:extrapolation}e), average extrapolation performance are not the highest among the others mass points.
    
    \begin{figure}[h!]
        \centering
        
        \begin{subfigure}{0.32\textwidth}
            \includegraphics[width=\textwidth]{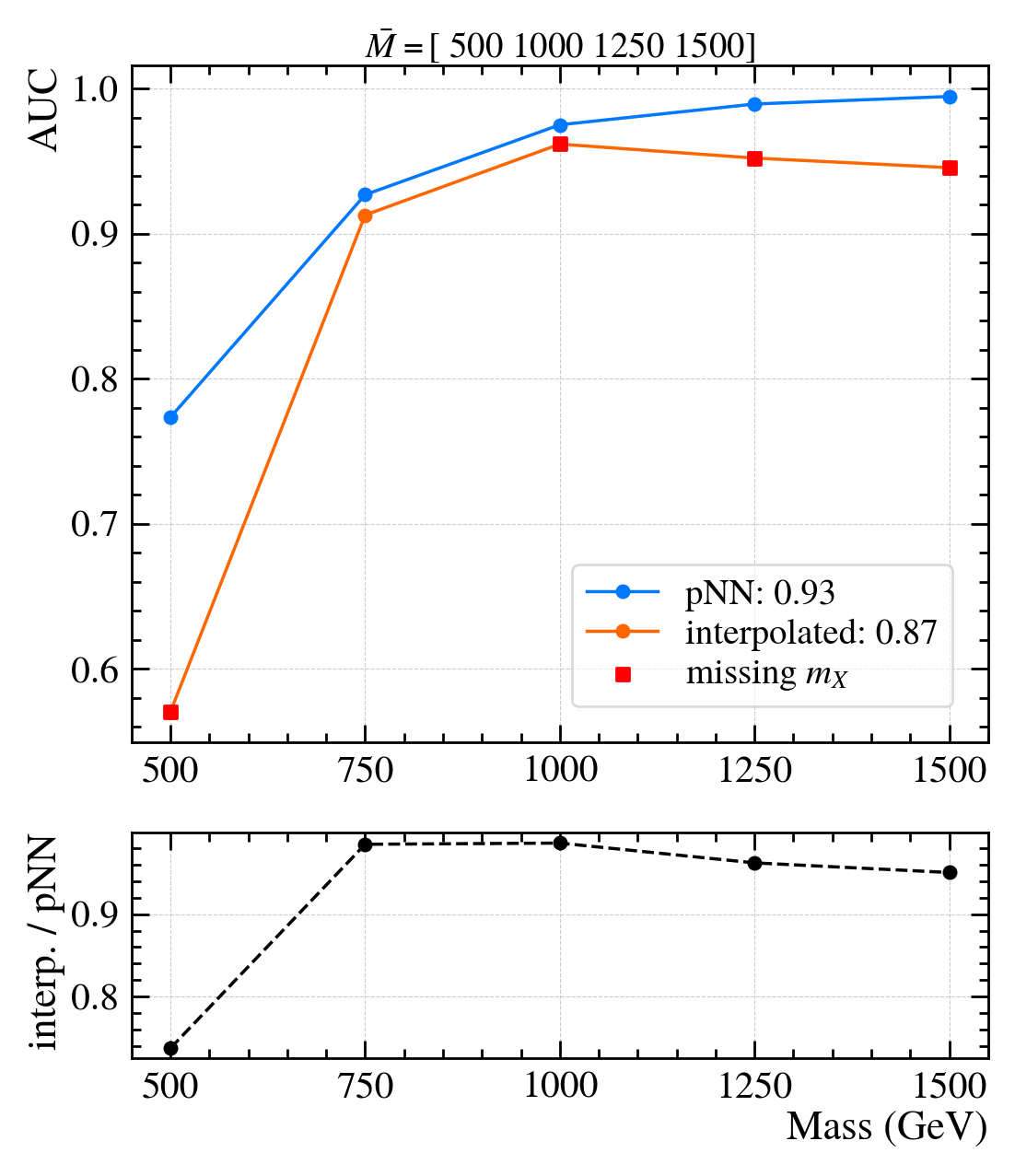}
            \caption{Trained on $\mathcal{D}_{750}$}
        \end{subfigure}
        \hfill
        \begin{subfigure}{0.32\textwidth}
            \includegraphics[width=\textwidth]{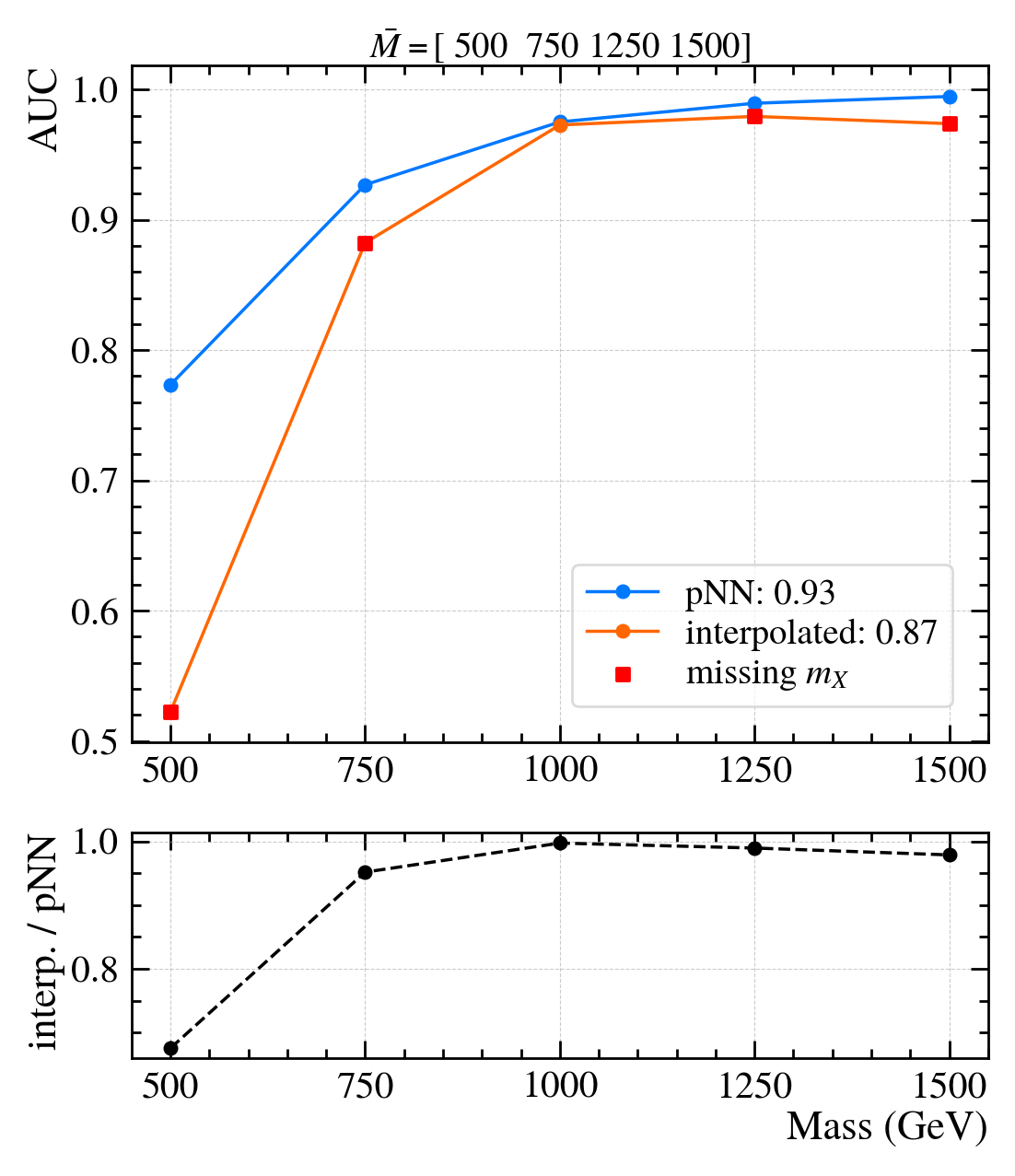}
            \caption{Trained on $\mathcal{D}_{1000}$}
        \end{subfigure}
        \hfill
        \begin{subfigure}{0.32\textwidth}
            \includegraphics[width=\textwidth]{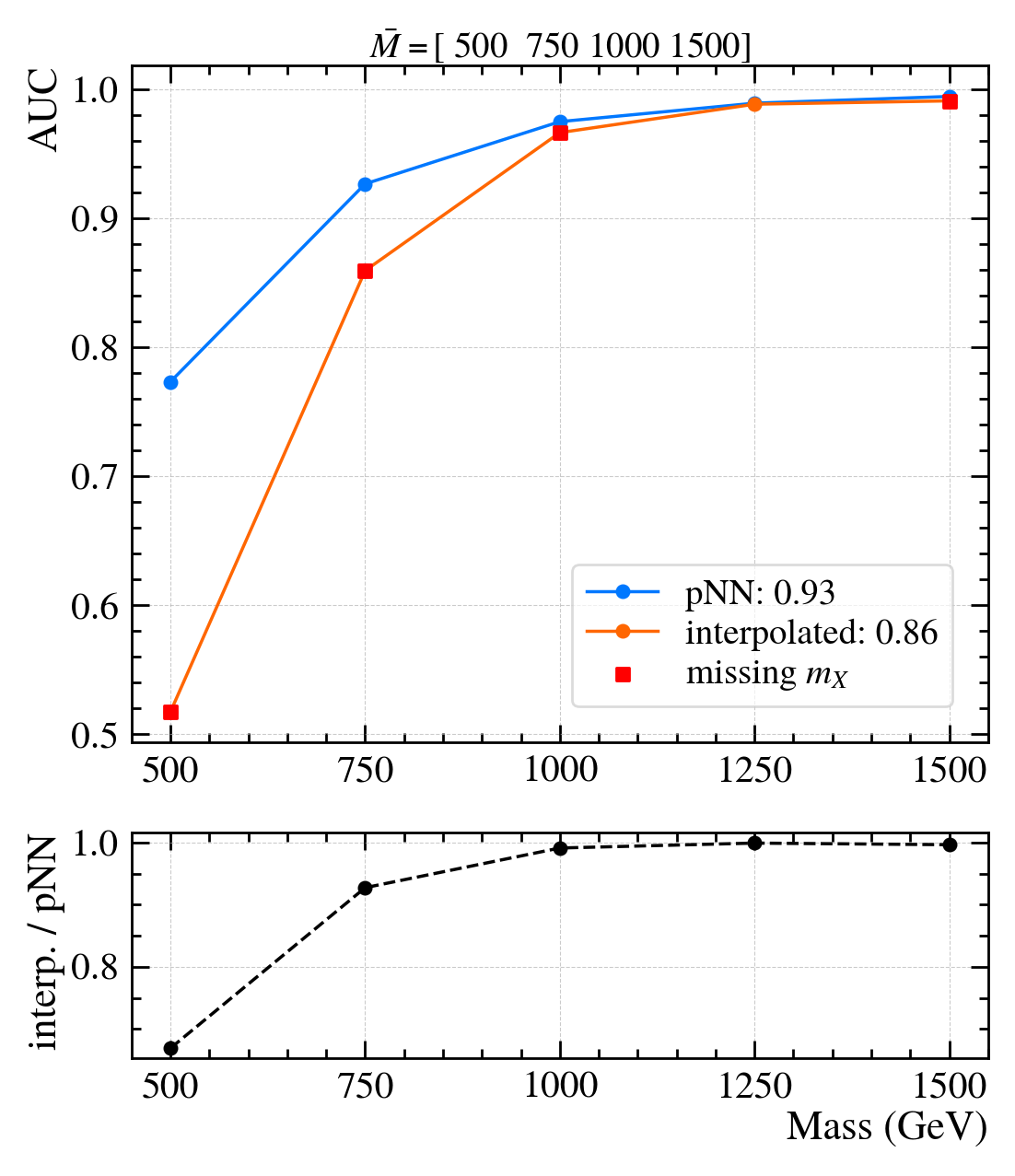}
            \caption{Trained on $\mathcal{D}_{1250}$}
        \end{subfigure}

        \begin{subfigure}{0.49\textwidth}
            \includegraphics[width=\textwidth]{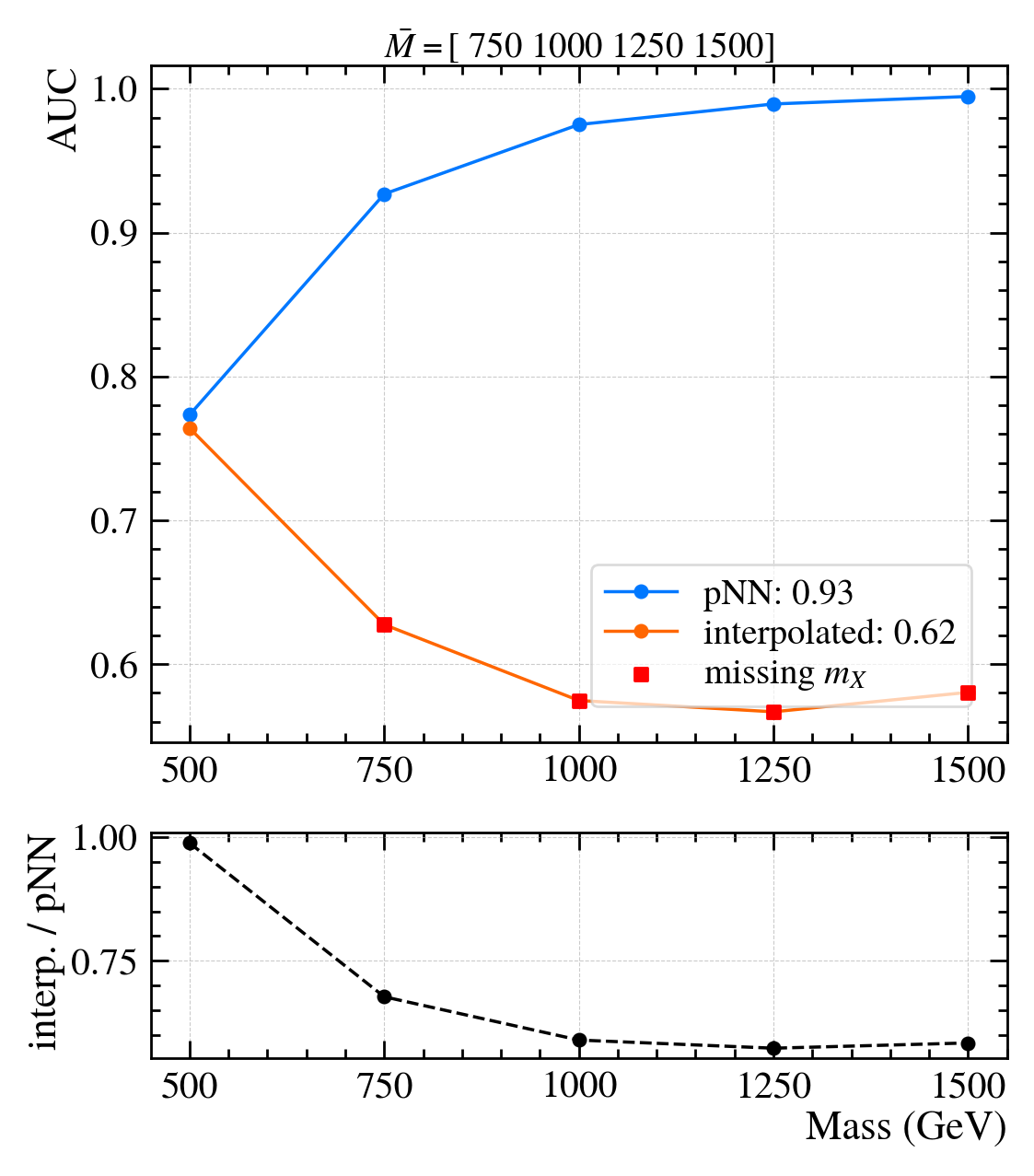}
            \caption{Trained on $\mathcal{D}_{500}$}
        \end{subfigure}
        \hfill
        \begin{subfigure}{0.49\textwidth}
            \includegraphics[width=\textwidth]{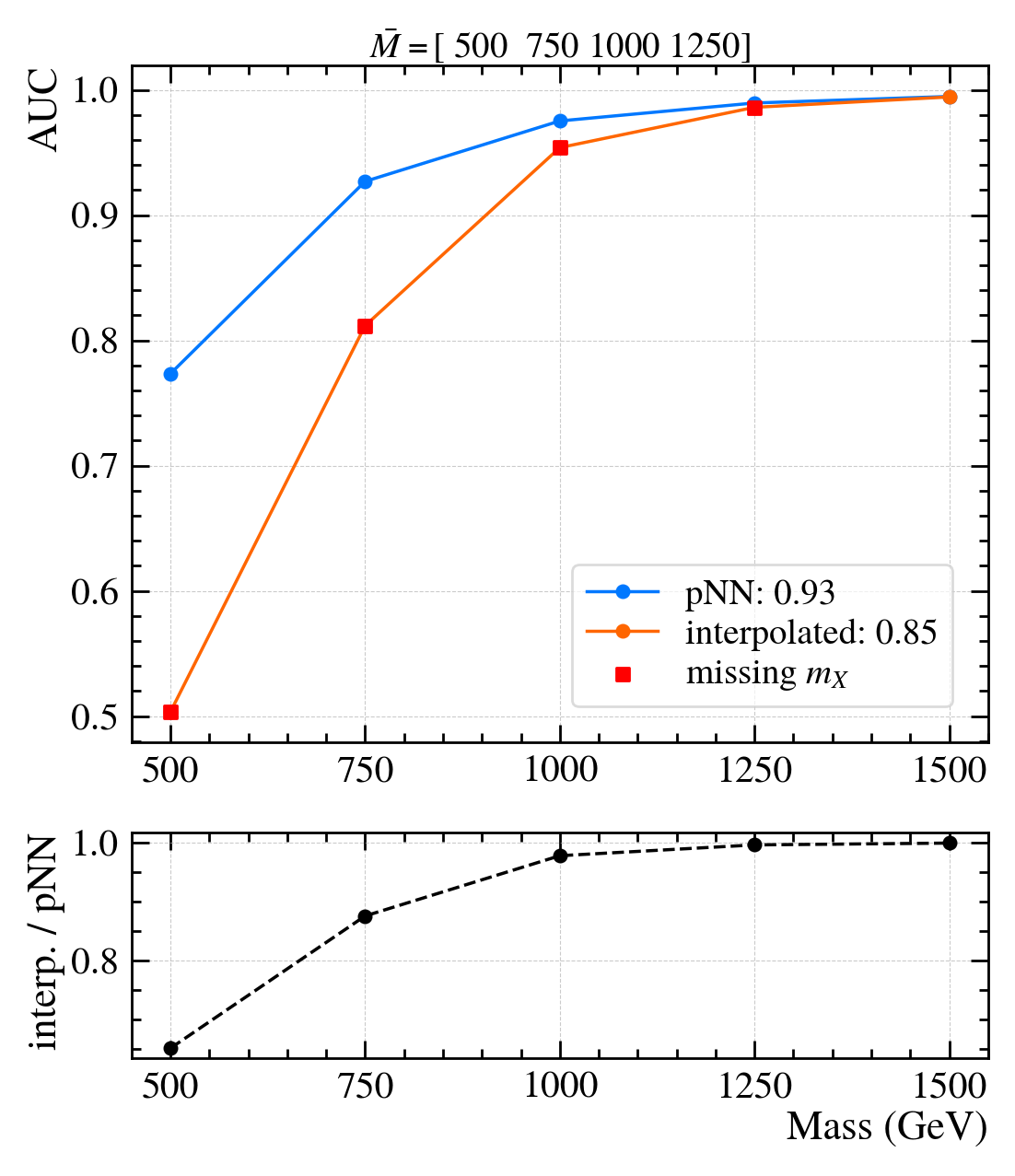}
            \caption{Trained on $\mathcal{D}_{1500}$}
        \end{subfigure}
        
        \caption{Extrapolation on \hepmass. We can clearly notice how different mass $500$ is from all the others, in fact its extrapolation performance is really poor: the (mean) AUC decreases by more than $30\%$. The other plots present a mean loss in AUC that is at most $8\%$.}
    \label{fig:extrapolation}
    \end{figure}
    
    \item \textbf{Background distribution and Regularization}: the impact of background distribution (section \ref{subsec:bkg}) goes beyond classification performance, as it may also affect interpolation. Figure \ref{fig:interp_failure_imb} denotes a pNN that hardly interpolates; such network was trained on a \textit{uniformly distributed} background, without regularization. 
    During training on all the mass points, we noticed that the same network were able to almost classify perfectly both the training and validation sets: clearly overfitting them. 
    As discussed previously, having a uniformly distributed mass feature for the background introduces an additional correlation with the class label, making training "easy". Indeed, by regularizing the model enough and increasing the batch size, generalization as well interpolation can be achieved with success.
\end{itemize}

\paragraph{Select mass hypotheses.} Another practical aspect to consider is how to select the mass hypotheses to drop for a fair measure of interpolation. As described earlier, the goodness of training data can mislead us when quantifying interpolation. So, we suggest to drop almost half of the mass hypotheses, in the following way: let's assume we have hypotheses $[m_0, m_1, m_2, m_3, m_4, m_5]$, we may drop $[m_0, m_2, m_4]$ or $[m_1, m_3]$. Sometimes, it is interesting (also useful) to see what a parametric network can achieve when trained on \textit{only one mass}, and evaluated for extrapolation on all the others: as shown in figure \ref{fig:extrapolation}. Such a test may resemble the training of individual classifiers, being different in that also the mass feature is provided. This kind of test can be useful to better understand the contribution of both network architecture and training data to the quality of the resulting extrapolation: in this case, we can expect the parametrized network (trained on only one mass) to perform well on the only training mass, but not too worse on immediately \textit{close} hypotheses also maintaining a reasonable accuracy on \textit{far} masses. This can be an easy way to asses the similarity of features among masses (which is an intrinsic property of the training data): if masses are similar, the network should perform almost the same on each unseen mass.

\newpage
\subsection{Learned Mass Representation}
\label{subsec:mass_representation}
\begin{figure}[h!]
    \centering
    \begin{subfigure}{0.495\textwidth}
        \includegraphics[width=\textwidth]{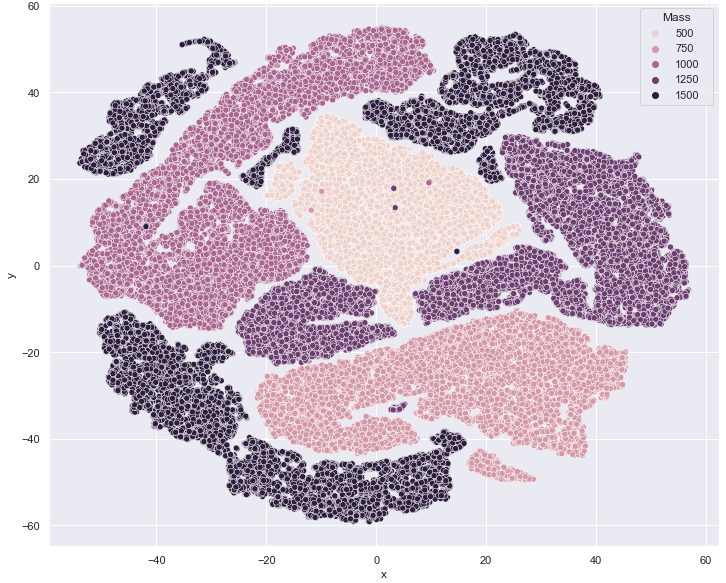}
        \caption{Mass label}
    \end{subfigure}
    \begin{subfigure}{0.495\textwidth}
        \includegraphics[width=\textwidth]{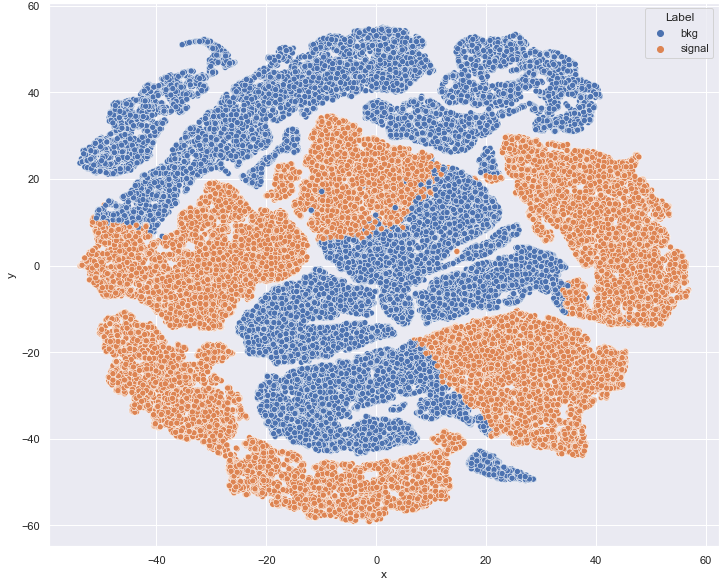}
        \caption{Class label}
    \end{subfigure}
    
    \caption{Visualization of the intermediate learned representation (i.e. last hidden layer) of each individual network by means of t-SNE \cite{van2008visualizing, wattenberg2016use} on the \hepmass\ dataset. By coloring the points by the mass label (left plot), we notice that representations related to the same mass are clustered together, meaning that each network $g_i$ has indirectly acquired knowledge about its underlying signal mass hypothesis $m_i$ (although not given as input). Also a structured part of them clearly depicts the learned class label (right plot).}
    \label{fig:tsne_individual}
\end{figure}

From section \ref{sec:intro}, in our signal-background classification problem, we actually consider $\mathcal M$ mass hypotheses for the signal. This means that our original task can be broken down into $|\mass|$ smaller classification problems, each of them considering only a specific mass $m_i\in \mathcal M$. In fact, approaches before the parametric network \cite{baldi2014searching} used to solve each sub-task by training a neural network solely on $\mathcal{D}_{m_i}$ (i.e. a slice of the original dataset $\mathcal D$, that selects events whose mass feature is $m_i$), thus obtaining a (disjoint) set of $\mathcal M$ individual networks, which we will call $g_{m_i}$ (or $g_i$ for short). Somehow each individual network $g_i$, despite being trained solely on one mass, is able to \textit{implicitly} relate the input features to the signal hypothesis $m_i$ they truly belong to (or even to the underlying invariant mass). This fact seems to be confirmed by the visualization in figure \ref{fig:tsne_individual}, in which each intermediate representation $h_i$ (of network $g_{m_i}$ for all $m_i\in \mathcal M$) has a precise and nicely clustered spatial arrangement, that also relates well to the learned class label.

\begin{figure}[h]
    \centering
    \begin{subfigure}{0.495\textwidth}
        \includegraphics[width=\textwidth]{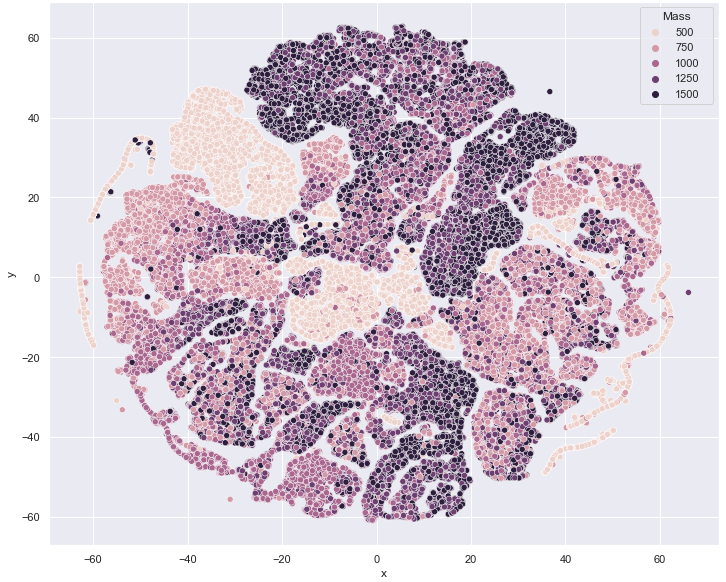}
        \caption{Mass label}
    \end{subfigure}
    \begin{subfigure}{0.495\textwidth}
        \includegraphics[width=\textwidth]{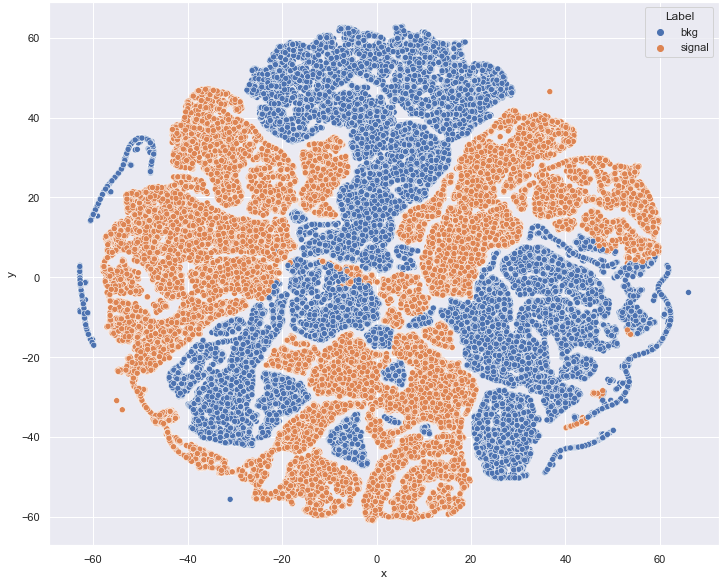}
        \caption{Class label}
    \end{subfigure}
    
    \caption{Visualization of the intermediate learned representation of a pNN trained on the \hepmass\ dataset. Compared to figure \ref{fig:tsne_individual}, some structure is present although more convoluted, and less clear in general.}
    \label{fig:tsne_pnn}
\end{figure}

Such kind of visualizations may provide further insights about the relation existent between a parametric network and a set of individual networks. Intuitively, we may want the intermediate representation of our pNN to be disentangled along the \textit{mass "axis"} (in the underlying manifold), as seen in figure \ref{fig:tsne_individual} for individually trained neural networks (considered as a whole). The situation for the pNN is similarly structured (figure \ref{fig:tsne_pnn}): some mass are well clustered (e.g. at $500$ GeV) and for others we can observe a smooth "shading" among them. This means that the pNN has \textit{partially} recovered the underlying structure about the individual masses, but there is still some confusion about representing datapoints coming from higher values of the signal mass hypotheses.

\section{Results}
\label{sec:evaluation}
Since the datasets we have for signal-background classification can be divided into $|\mass|$ groups (in order to be able to "parametrize" a neural network), also evaluation metrics have to be considered in terms of the available mass hypotheses for the signal. In particular, the models are evaluated on each $m_i$ separately. So we consider the signal events generated at a certain $m_i\in \mass$, along with the whole background: i.e. the background events that spans the entire mass range. Indeed, the results provided in this section were all computed only on the test-set of the respective datasets. Moreover, we \textit{weight} both signal and background samples (only for evaluation) such that the weighted count of signal events is equal to the weighted count of background events, i.e:
\begin{equation}
    \sum_i w_s^{(i)} = \sum_j w_b^{(j)}
\end{equation}
In particular, for both datasets we set the signal weight to one, $w_s=1$, and for the background as $w_b = 1/5$; since we have five mass hypotheses for the signal. For such reason, each time we test a particular mass hypothesis $m_i\in \mass$, we select the corresponding signal (i.e. all the signal events that have $m_i$ as mass feature) and the \textit{whole} background: i.e. the mass feature for all background samples $b$, is set equal to $m_i$; thus, $m^{(b)} = m_i$. This is to account for the fact that, in \hepmass, the original background's mass is assigned randomly, being sampled from the set $\mass$ (i.e. the \textit{identical fixed} strategy, discussed in section \ref{subsec:bkg}). 

\subsection{Metrics}
We consider standard evaluation metrics for classification tasks, such as the \textit{AUC} (area under the curve) of the \textit{ROC} (receiving operating characteristic) and \textit{Precision-Recall} curves. In particular, the ROC curve can be interpreted for HEP as comparing the \textit{signal efficiency} ($y$-axis) against the \textit{background efficiency} ($x$-axis): in terms of how much signal is retained when considering a certain fraction of the background. Otherwise, we can consider the \textit{background rejection} (i.e. $1-\text{background efficiency}$): how much signal is retained at a certain discard of background. The Precision-Recall curve instead, compares the signal efficiency (\textit{recall}) with what we call the \textit{purity} (precision): the number of true signal divided by the number of events classified as signal (which also contains misclassification of the background). 

Along usual classification metrics, we also consider the \textit{Approximate Median Significance} (AMS) \cite{adam2015higgs, cowan2011asymptotic} but in the following form:
\begin{equation}
    \text{AMS}(t) = \frac{s_t}{\sqrt{s_t + b_t}},
\end{equation}
in which $s_t$ and $b_t$ is the (weighted\footnote{In general, the weights $w_s$ and $w_b$ are introduced with the only aim of balancing the occurrences of the two classes when testing for a particular $m_i$: they have no physical meaning, as we want to keep our study as general as possible, without assuming any luminosity and signal cross section weights during training and evaluation.}) number of \textit{true} signal and true background events, respectively, that passed the classification threshold $t$. This quantity is useful to determine an optimal classification threshold for our networks, called the \textit{best cut} $t^\star$, that is the threshold that maximizes the significance: $t^\star = \arg\max_t \text{AMS}(t)$. Since the value of the AMS depends on the number of events from which is calculated, we propose a new metric the \textit{significance ratio} ($\sigma_\text{ratio}$) that is normalized in $[0, 1]$, thus being very intuitive to interpret. The significance ratio is defined as the ratio between the best (maximum) AMS by the largest possible significance (only achievable ideally, by means of a perfect classification when $s_t=s_{\max}$, i.e. equal to all the true signal, and $b_t=0$):
\begin{equation}
    \label{eq:ams_ratio}
    \sigma_\text{ratio} = \frac{\max_t \text{AMS}(t)}{s_{\max} / \sqrt{s_{\max}}} = \max_t \bigg\{ \frac{s_t\cdot \sqrt{s_{\max}}}{s_{\max}\cdot\sqrt{s_t + b_t}} \bigg\} = \frac{s_\star\cdot\sqrt{s_{\max}}}{s_{\max}\cdot\sqrt{s_\star+b_\star}},
\end{equation}
where $s_\star = s_{t^\star}$, and $b_\star = b_{t^\star}$. Such metric can be also used to compare how well the same model classifies different mass hypotheses: this is now possible since the number of events belonging to a certain $m_i$ does not affect the scale of the metric (as happens for the regular AMS, instead), anymore. We can further say that the best cut $t^\star$, apart from telling us which classification threshold is the best to determine the positive class, is also an useful quantity to monitor because it can provide additional information about the goodness of the classification. In particular, by plotting the best cut versus the mass we may observe failure cases in which $t^\star$ is either $0$ or $1$, depicting a situation in which the network is unable to correctly separate out the background ($t^\star = 0$) or to retain a significant amount of signal ($t^\star = 1$). A special failure case can be observed when $s_\star=s_{\max}$ and $s_\star=b_\star$, i.e. the signal is equal in number to all the true signal, which is also equal to the true classified background (e.g. due to applied weights). In this case we would obtain $\sigma_\text{ratio} = 1/\sqrt{2}\approx 0.707$, since:
\begin{equation}
    \frac{s_\star\cdot \sqrt{s_{\max}}}{s_{\max}\cdot \sqrt{s_\star+b_\star}} = \frac{\sqrt{s_\star}}{\sqrt{s_\star + s_\star}} = \frac{1}{\sqrt{2}}
\end{equation}

Indeed, measuring $\sigma_\text{ratio}\approx 0.707$ also implies having an AUC of $0.5$, corresponding to nonsense classification. Moreover, if the best cut $t^\star$ is such that $s_\star = s_{\max}$ but $b_\star > 0$, then $\sigma_\text{ratio} = \frac{\sqrt{s_\star}}{\sqrt{s_\star + b_\star}}$ decreases toward zero (in the limit), as $b_\star$ approaches $b_{\max}$ (i.e. the weighted count of all true background events).

\subsection{Baselines}
 To first assess the advantages brought by parametric neural networks we should compare them to their "non-parametric" counterparts, namely \textit{single} and \textit{individual} neural networks. What we call a single-NN is just a neural network that is trained \textit{without} the mass feature ($m$) as input but on all $\mass$ hypotheses at the same time; so, it has only one input: the features $x$. Instead, the individual networks are, as the name suggests, a set of single networks, $g_{\phi_i}$, each of them trained to target a specific mass hypothesis $m_i \in \mass$. Since each $g_{\phi_i}$ is trained in isolation on one $m_i$ against the whole background, we expect the individual networks to easily beat the single network as it should face a harder learning problem. Moreover, both kind of networks can provide baseline performance for interpolation: the single network is trained to fit all data regardless of the mass (not given as input), whereas each individual network $g_{\phi_i}$ will interpolate by means of the \textit{similarity} between the hypothesis $m_i$ it was trained on, and the $m_j$ to interpolate.
 
 \begin{figure}
     \centering
     \includegraphics[width=\textwidth]{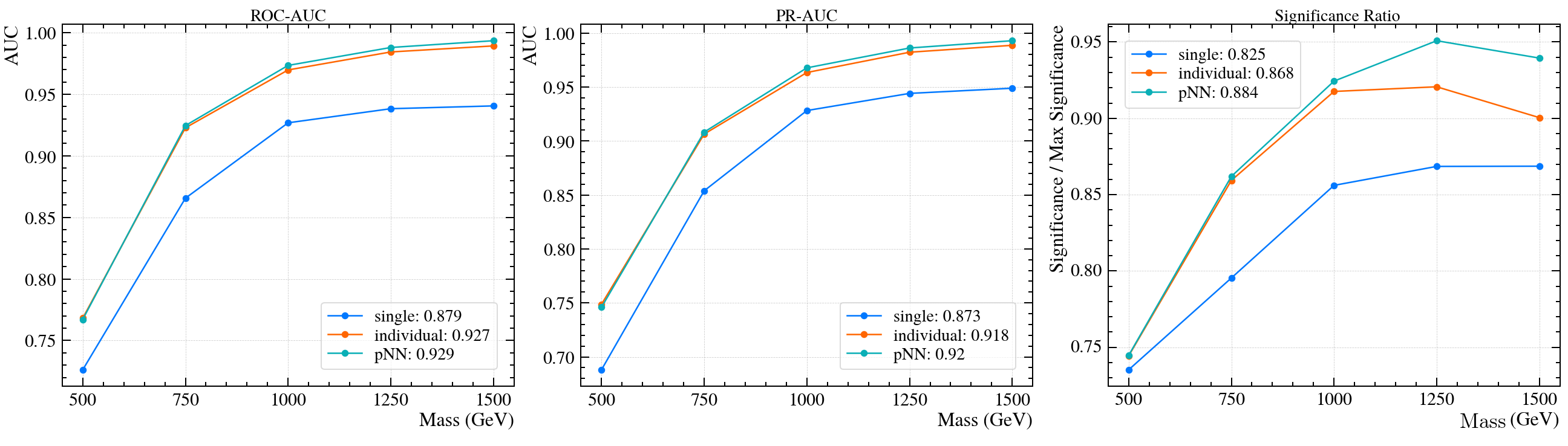}
     \caption{Comparison among (tuned) baselines on \hepimb. As we can see, the best classification performance (in terms of AUC of both ROC and PR curves, and $\sigma_\text{ratio}$ as well; for each $m_i\in\mass$) are achieved by the individual NNs, as well as the parametric baseline (with identical-fixed distribution for the background's mass feature).}
     \label{fig:imb_baselines}
 \end{figure}
 
 Finally, to assess the effectiveness of the various decision choices described in section \ref{sec:design_choices}, we apply them (whether possible) to the non-parametric baselines, and also (of course) to the parametric baseline: a \textit{vanilla pNN}, as intended by Baldi et al \cite{baldi2016parameterized}. Results for classification performance are shown in figure \ref{fig:imb_baselines}.

\subsection{Evaluation}
\label{subsec:eval}
\paragraph{Hyperparameters.} All the neural networks used for comparison are built using the TensorFlow 2.X \cite{abadi2016tensorflow} framework along with the Keras \cite{chollet2015keras} library for Python. To improve reproduce our results we fix the \textit{random seed} to be $42$. All the networks use the same hyperparameters: ReLU activation, $[300, 150, 100, 50$] units for each hidden layer, sigmoid output, binary-crossentropy loss, Adam optimizer \cite{kingma2014adam}, batch size of $1024$ (except when told otherwise), and default initialization: \texttt{glorot\_uniform} \cite{glorot2010understanding} for weights, and constant \texttt{zero} initializer for biases. It results in about $70$k learnable parameters. 
The learning rate is never decayed, and set to $3\times10^{-4}$. In general, we always use regularization by means of both dropout (with drop rate of $25\%$), and $l2$-weight decay. The weight decay is applied differently, with a strength of $10^{-4}$ for weights (or $10^{-5}$), and $10^{-5}$ for biases (or $10^{-6}$). Also, the same hyperparameters are kept for both datasets, as well as the training budget fixed at $25$ epochs. In general, the hyperparameters we use were initially tuned for the vanilla pNN architecture on \hepmass.

\begin{figure}[h]
    \centering
    \begin{subfigure}{0.32\textwidth}
        \includegraphics[width=\textwidth]{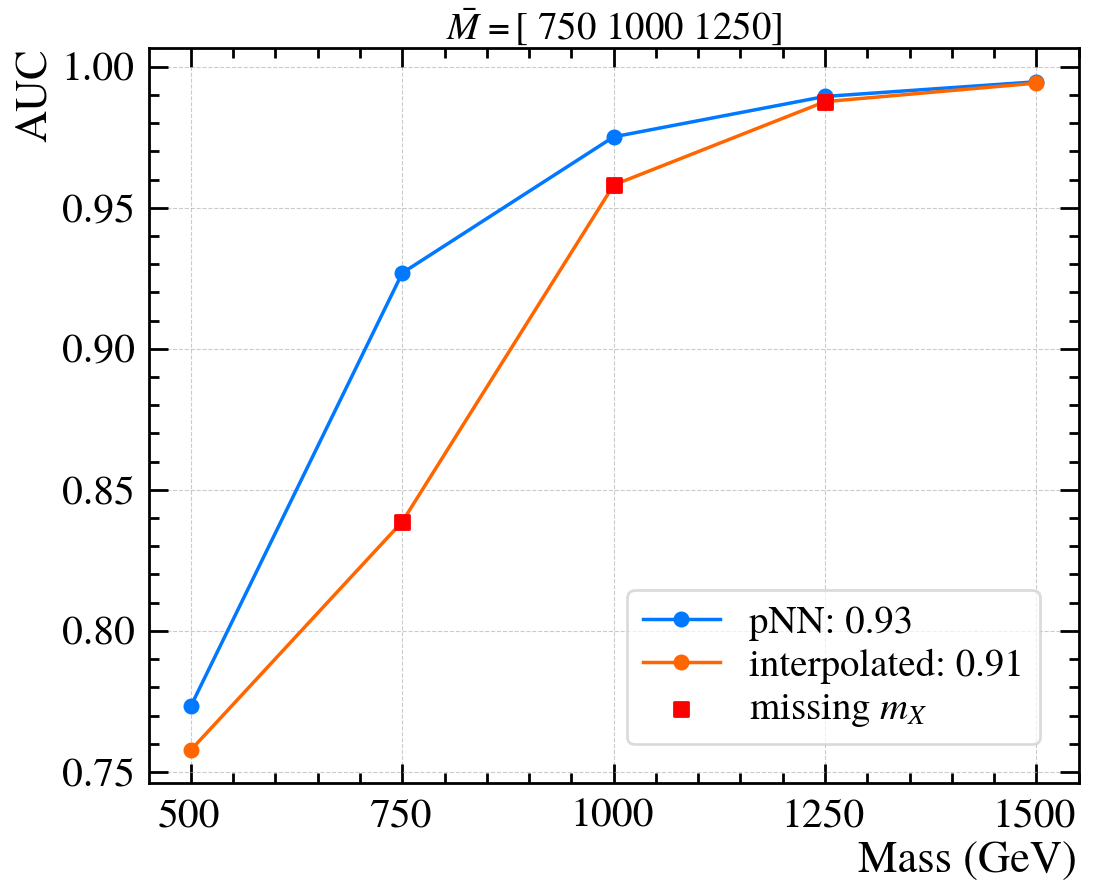}
        \caption{Interpolation on \hepmass.}
        \label{subfig:interp_hep1}
    \end{subfigure}
    \hfill
    \begin{subfigure}{0.32\textwidth}
        \includegraphics[width=\textwidth]{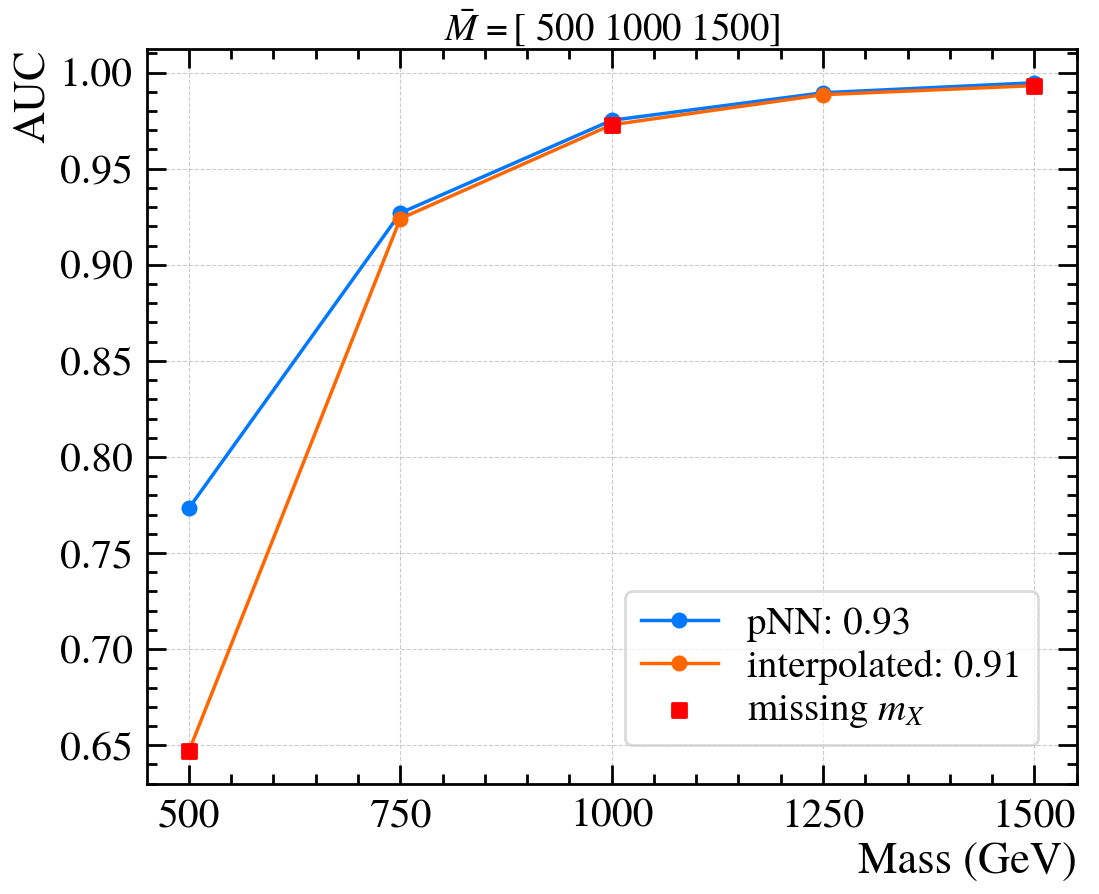}
        \caption{Interpolation on \hepmass.}
        \label{subfig:interp_hep2}
    \end{subfigure}
    \hfill
    \begin{subfigure}{0.32\textwidth}
        \includegraphics[width=\textwidth]{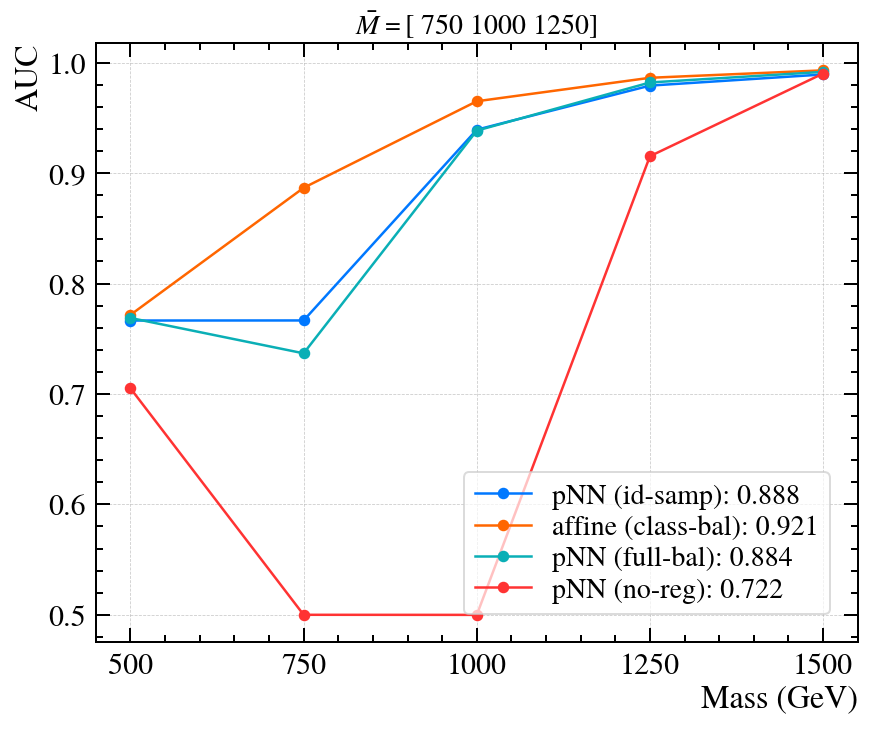}
        \caption{Interpolation on \hepimb.}
        \label{fig:interp_failure_imb}
    \end{subfigure}
    \caption{In figure (a) and (c) the models have been trained solely on two mass hypotheses: at $500$ and $1500$ GeV. Instead, the networks depicted in plot (b), were trained on mass $750$ and $1250$ GeV. The title of each plot denotes the missing masses by $\bar{M}$. In particular, in figure (c) we can seen how the lack of proper regularization prevents the network (depicted by the red line) to interpolate the signal at the missing mass hypotheses. Moreover, the affine network outperforms by a large margin the two competent pNNs.}
\label{fig:interp_results}
\end{figure}

\paragraph{HEPMASS.} Results about classification performance (with baseline models), background's mass feature distribution, and model architecture are presented in table \ref{tab:result_auc_hepmass}. Whereas further results for interpolation are showed in figures \ref{subfig:interp_hep1} and \ref{subfig:interp_hep2}. 

\begingroup
    \setlength{\tabcolsep}{3pt}  
    \renewcommand{\arraystretch}{1.5}
    
    \begin{table}[h]
        \centering
        
        \begin{tabular}{cccccccc}
            \toprule
            \multicolumn{2}{c}{\textbf{Model}} & \multicolumn{5}{c}{\textbf{Mass (GeV)}} & \multirow{2}{*}{\textbf{AUC (average)}} \\ 
            Kind & Mass Distribution & \multicolumn{1}{c}{500} & \multicolumn{1}{c}{750} & \multicolumn{1}{c}{1000} & \multicolumn{1}{c}{1250} & \multicolumn{1}{c}{1500} &  \\
            \midrule
            Single-NN & None & 68.55 & 89.37 & 96.38 & 98.07 & 98.69 & 90.21\% \\
            \textbf{Individual-NNs} & None & \textbf{77.35} & \textbf{92.59} & \textbf{97.42} & \textbf{98.89} & \textbf{99.45} & \textbf{93.14\%} \\
            pNN (linear) & identical (\textit{fixed}) & 63.42 & 89.36 & 96.08 & 97.90 & 98.58 & 89.07\% \\
            \hline
            pNN & uniform (\textit{fixed}) & 70.95 & 91.80 & 97.26 & 98.79 & 99.37 & 91.63\% \\
            Affine & uniform (\textit{fixed}) & 71.24 & 90.81 & 96.94 & 98.64 & 99.26 & 91.38\% \\
            pNN & uniform (\textit{sampled}) & 71.63 & 91.71 & 97.25 & 98.82 & 99.39 & 91.76\% \\
            Affine & uniform (\textit{sampled}) & 70.16 & 91.61 & 97.16 & 98.75 & 99.34 & 91.40\% \\
            pNN & identical (\textit{fixed}) & 76.78 & 92.56 & 97.46 & 98.92 & 99.46 & 93.04\% \\
            \textbf{Affine} & \textbf{identical (\textit{fixed})} & \textbf{77.34} & \textbf{92.80} & \textbf{97.55} & \textbf{98.96} & \textbf{99.49} & \textbf{93.23\%} \\
            pNN & identical (\textit{sampled}) & 76.77 & 92.60 & 97.48 & 98.92 & 99.46 & 93.05\% \\
            \textbf{Affine} & \textbf{identical (\textit{sampled})} & \textbf{77.31} & \textbf{92.77} & \textbf{97.55} & \textbf{98.96} & \textbf{99.49} & \textbf{93.22\%} \\
            \bottomrule
        \end{tabular}
        \caption{Per-mass ROC's AUC metric computed on the test-set of \hepmass: all AUC values are in percentage, higher is better. Best results are shown boldface. Options for mass distribution are described in section \ref{subsec:bkg}. We have also included a third baseline, a parametric network with linear activation: as we can see it underperforms even the single-NN, despite leveraging the additional information provided by the input mass feature.}
        \label{tab:result_auc_hepmass}
    \end{table}
\endgroup

\begin{figure}[h]
    \centering
    
    \begin{subfigure}{1\textwidth}
        \includegraphics[width=\textwidth]{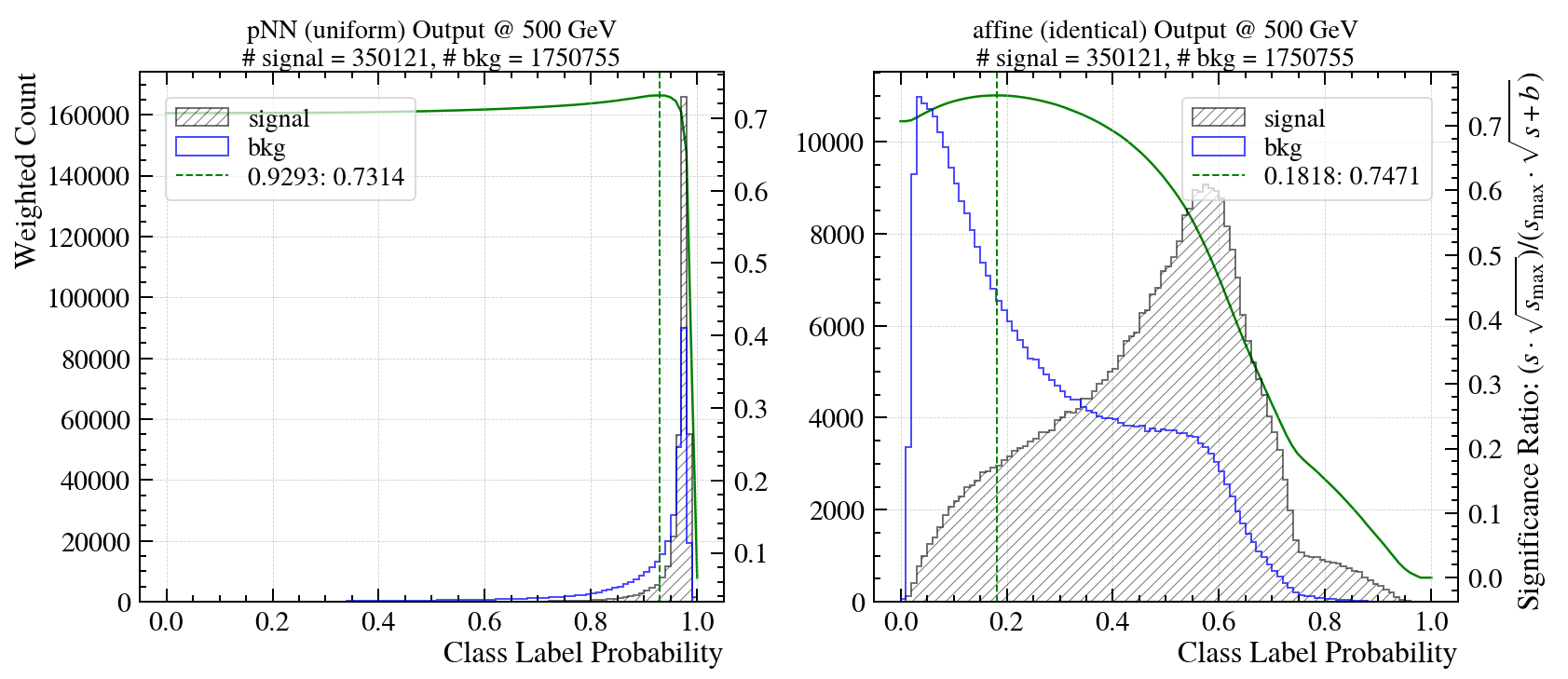}
        \caption{Weighted class separation (histograms), significance ratio (solid green curve), and best-cut (vertical dashed green line), at $m_X = 500$ GeV on \hepimb. The value of the best cut (classification threshold), and the corresponding value of the significance ratio are shown in the legend at the top.}
    \end{subfigure}
    \begin{subfigure}{1\textwidth}
        \includegraphics[width=\textwidth]{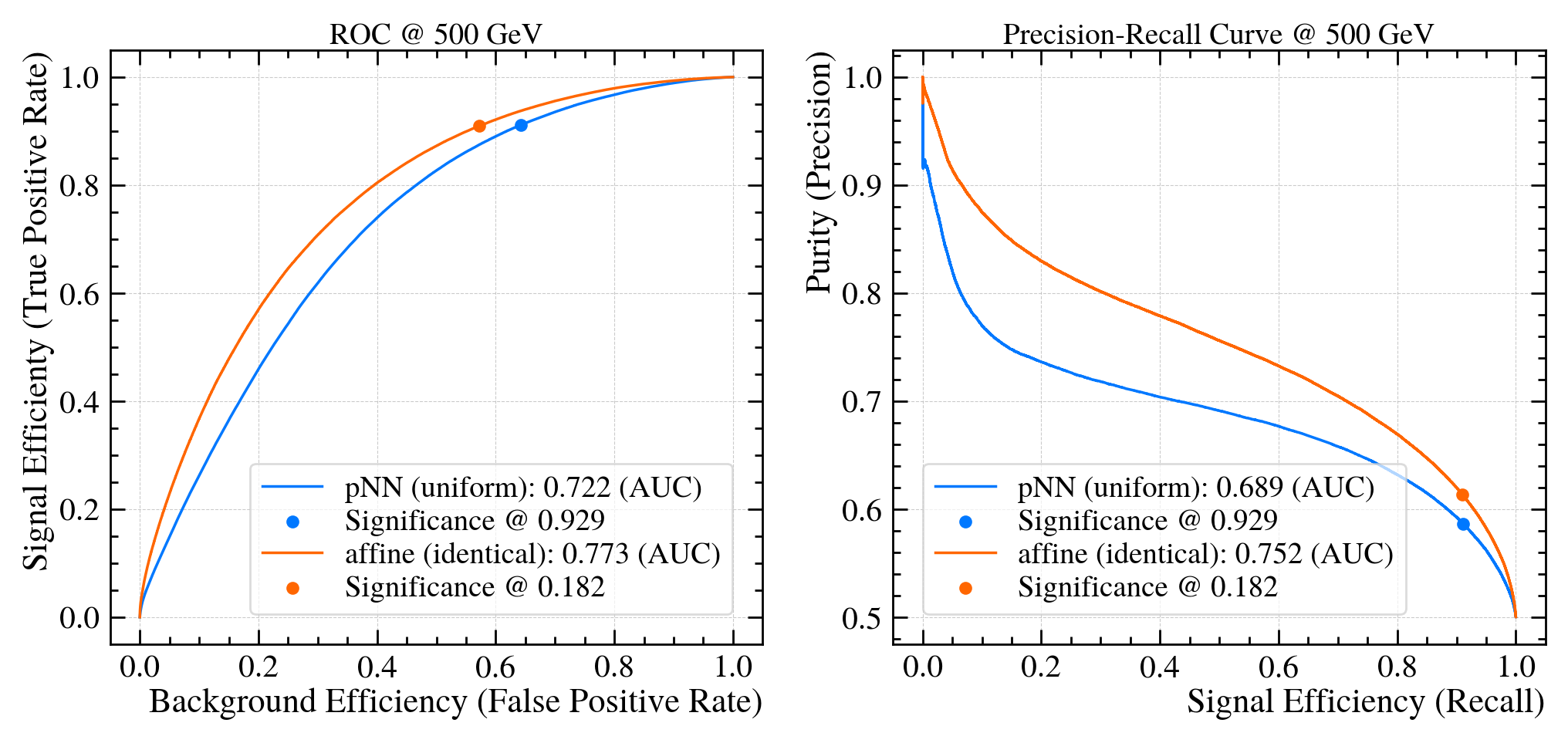}
        \caption{Comparison of ROC and PR curves at $m_X = 500$ GeV.}
    \end{subfigure}
    \caption{Comparison of classification performance on \hepimb, between a pNN (\textit{uniform}, sampled) and affine model (\textit{identical}, sampled). We can notice how the two models opt for rather different classification thresholds.}
\label{fig:imb_classification_results}
\end{figure}

\paragraph{HEPMASS-IMB.} Results on interpolation are presented both in table \ref{tab:result_hepimb_interp} and figure \ref{fig:interp_failure_imb}. Furthermore, an exhaustive comparison among baseline models, parametric and affine architectures, background's mass distribution, and training procedure is detailed in tables \ref{tab:result_hepimb} and \ref{tab:result_hepimb2}. Finally, outcomes about classification performance in terms of class separation are detailed in figure \ref{fig:imb_classification_results}.

\begingroup
    \setlength{\tabcolsep}{6pt}
    \renewcommand{\arraystretch}{1.5}
    
    \begin{table}[h]
        \centering
    
        \begin{tabular}{ccccccccc}
            \toprule
            \multicolumn{2}{c}{\textbf{Model}} & \multicolumn{5}{c}{\textbf{Mass (GeV)}} & \multicolumn{2}{c}{\textbf{Average (\%)}} \\
            Kind & Mass Distribution & 500 & 750 & 1000 & 1250 & 1500 & AUC & $\sigma_\text{ratio}$ \\
            \midrule
            pNN & identical (\textit{sampled}) & 76.65 & 76.66 & 93.93 & 97.92 & 98.93 & 88.82 &  \\
             &  & 74.38 & 74.78 & 88.38 & 93.40 & 91.52 & & 84.49 \\
            pNN (class) & identical (\textit{sampled}) & 77.41 & 71.25 & 94.60 & 98.48 & 99.29 & 88.21 & \\
            &  & 74.81 & 73.23 & 89.16 & 94.55 & 96.32 & & 85.61 \\
            pNN (full) & identical (\textit{sampled}) & 76.88 & 73.68 & 93.85 & 98.21 & 99.17 & 88.36 & \\
             &  & 74.56 & 75.38 & 88.39 & 93.67 & 93.82 & & 85.16 \\
            \hline
            Affine & identical (\textit{sampled}) & 76.49 & 88.79 & 96.39 & 98.44 & 99.13 & 91.85 & \\
             &  & 74.33 & 82.66 & 91.07 & 93.86 & 89.90 & & 86.36 \\
            \textbf{Affine (class)} & \textbf{identical (\textit{sampled})} & 77.16 & 88.67 & 96.51 & 98.63 & 99.31 & \textbf{92.05} \\
            & & 74.70 & 83.10 & 91.30 & 94.80 & 96.43 & & \textbf{88.07} \\
            Affine (full) & identical (\textit{sampled}) & 76.20 & 84.68 & 94.87 & 97.87 & 98.89 & 90.50 \\
            & & 74.26 & 80.28 & 89.04 & 92.91 & 89.84 & & 85.27 \\
            \bottomrule
        \end{tabular}
        \caption{Per-mass and averaged, ROC's AUC and significance ratio (eq. \ref{eq:ams_ratio}) metrics computed on \hepimb, showing interpolation capabilities. The missing signal mass hypotheses are $\bar\mass = \{750, 1000, 1250\}$. Best results are shown in boldface. The \textit{identical (sampled)} distribution strategy is described in section \ref{subsec:bkg}, and the \textit{balanced training} is detailed in section \ref{subsec:balanced_training}.}
        \label{tab:result_hepimb_interp}
    \end{table}
\endgroup

\begingroup
    \setlength{\tabcolsep}{6pt}
    \renewcommand{\arraystretch}{1.5}
    
    \begin{table}[h]
        \centering
    
        \begin{tabular}{ccccccccc}
            \toprule
            \multicolumn{2}{c}{\textbf{Model}} & \multicolumn{5}{c}{\textbf{Mass (GeV)}} & \multicolumn{2}{c}{\textbf{Average (\%)}} \\
            Kind & Mass Distribution & 500 & 750 & 1000 & 1250 & 1500 & AUC & $\sigma_\text{ratio}$ \\
            \midrule
            Single-NN & None & 72.62 & 86.56 & 92.68 & 93.83 & 94.06 & 87.95 &  \\
             &  & 73.52 & 79.51 & 85.59 & 86.85 & 86.86 & & 82.47 \\
            Individual-NNs & None & 76.81 & 92.28 & 96.98 & 98.44 & 98.93 & 92.69 &  \\
             &  & 74.43 & 85.91 & 91.75 & 92.07 & 90.04 & & 86.84 \\
             \hline
            pNN & identical (\textit{fixed}) & 76.71 & 92.46 & 97.35 & 98.80 & 99.35 & 92.93 &  \\
             &  & 74.46 & 86.17 & 92.43 & 95.09 & 93.94 & & 88.42 \\
            pNN (class) & identical (\textit{fixed}) & 76.93 & 92.56 & 97.41 & 98.85 & 99.39 & 93.03 & \\
             &  & 74.67 & 86.34 & 92.55 & 95.32 & 96.64 & & 89.11 \\
            pNN (full) & identical (\textit{fixed}) & 76.29 & 92.33 & 97.33 & 98.79 & 99.34 & 92.82 &  \\
             &  & 74.31 & 86.14 & 92.44 & 95.17 & 96.47 & & 88.91 \\
            Affine & identical (\textit{fixed}) & 77.19 & 92.62 & 97.43 & 98.86 & 99.40 & 93.10 \\
            & & 74.70 & 86.30 & 92.52 & 95.31 & 95.61 & &  88.89 \\
            \textbf{Affine (class)} & \textbf{identical (\textit{fixed})} & 77.19 & 92.64 & 97.46 & 98.88 & 99.42 & \textbf{93.12} & \\
            & & 74.76 & 86.38 & 92.62 & 95.37 & 96.77 & & \textbf{89.18} \\
            Affine (full) & identical (\textit{fixed}) & 76.45 & 92.46 & 97.38 & 98.78 & 99.31 & 92.88 \\
            & & 74.42 & 86.24 & 92.50 & 95.07 & 95.07 & & 88.66 \\
            \hline
            pNN & identical (\textit{sampled}) & 76.73 & 92.44 & 97.34 & 98.80 & 99.35 & 92.93 \\
            & & 74.47 & 86.11 & 92.37 & 94.96 & 93.76 & & 88.34 \\
            pNN (class) & identical (\textit{sampled}) & 76.88 & 92.51 & 97.40 & 98.86 & 99.40 & 93.01 \\
            & & 74.61 & 86.29 & 92.51 & 95.33 & 96.70 & & 89.09 \\\
            pNN (full) & identical (\textit{sampled}) & 76.38 & 92.39 & 97.34 & 98.79 & 99.34 & 92.85 \\
            & & 74.35 & 86.17 & 92.47 & 95.16 & 96.52 & & 88.93 \\
            Affine & identical (\textit{sampled}) & 77.26 & 92.64 & 97.43 & 98.87 & 99.40 & 93.12 \\
            & & 74.71 & 86.34 & 92.55 & 95.33 & 95.74 & & 88.93 \\
            \textbf{Affine (class)} & \textbf{identical (\textit{sampled})} & 77.24 & 92.66 & 97.46 & 98.88 & 99.42 & \textbf{93.13} \\
            & & 74.78 & 86.39 & 92.60 & 95.39 & 96.77 & & \textbf{89.19} \\
            Affine (full) & identical (\textit{sampled}) & 76.41 & 92.43 & 97.37 & 98.80 & 99.34 & 92.87 \\
            & & 74.37 & 86.19 & 92.47 & 95.10 & 95.23 & & 88.67 \\
            \bottomrule
        \end{tabular}
        \caption{Comparison of baselines, model architecture, background's mass feature distribution, and training procedure on \hepimb. Performances are evaluated in terms of ROC-AUC and significance ratio. Options for mass distribution are described in section \ref{subsec:bkg}. The words \textit{class} and \textit{full}, refer, respectively, to class- and fully-balanced training (section \ref{subsec:balanced_training}). Best results are shown boldface.}
        \label{tab:result_hepimb}
    \end{table}
\endgroup

\begingroup
    \setlength{\tabcolsep}{6pt}
    \renewcommand{\arraystretch}{1.5}
    
    \begin{table}[h]
        \centering
    
        \begin{tabular}{ccccccccc}
            \toprule
            \multicolumn{2}{c}{\textbf{Model}} & \multicolumn{5}{c}{\textbf{Mass (GeV)}} & \multicolumn{2}{c}{\textbf{Average (\%)}} \\
            Kind & Mass Distribution & 500 & 750 & 1000 & 1250 & 1500 & AUC & $\sigma_\text{ratio}$ \\
            \midrule
            pNN & uniform (\textit{fixed}) & 76.67 & 92.41 & 97.33 & 98.79 & 99.34 & 92.91 \\
            & & 74.48 & 86.15 & 92.39 & 95.15 & 94.70 & & 88.57 \\
            pNN (class) & uniform (\textit{fixed}) & 71.00 & 91.56 & 97.14 & 98.75 & 99.32 & 91.55 \\
            & & 72.71 & 85.64 & 92.22 & 95.09 & 96.45 & & 88.42 \\
            pNN (full) & uniform (\textit{fixed}) & 72.19 & 91.41 & 97.11 & 98.70 & 99.25 & 91.73 \\
            & & 73.06 & 85.42 & 92.16 & 94.93 & 96.23 & & 88.36 \\
            Affine & uniform (\textit{fixed}) & 77.26 & 92.62 & 97.43 & 98.86 & 99.41 & 93.11 \\
            & & 74.73 & 86.31 & 92.52 & 95.25 & 94.65 & & 88.69 \\
            \textbf{Affine (class)} & \textbf{uniform (\textit{fixed})} & 77.29 & 92.68 & 97.47 & 98.89 & 99.43 & \textbf{93.15} \\
            & & 74.82 & 86.42 & 92.64 & 95.40 & 96.80 & & \textbf{89.22} \\
            Affine (full) & uniform (\textit{fixed}) & 66.67 & 91.15 & 97.08 & 98.70 & 99.30 & 90.58 \\
            & & 70.75 & 85.46 & 92.05 & 94.88 & 96.34 & & 87.90 \\
            \hline
            pNN & uniform (\textit{sampled}) & 72.21 & 91.48 & 97.17 & 98.72 & 99.29 & 91.77 \\
            & & 73.14 & 85.55 & 92.15 & 94.88 & 95.17 & & 88.18 \\
            pNN (class) & uniform (\textit{sampled}) & 71.48 & 91.68 & 97.16 & 98.75 & 99.32 & 91.68 \\
            & & 72.80 & 85.61 & 92.20 & 95.10 & 96.47 & & 88.44\\
            pNN (full) & uniform (\textit{sampled}) & 71.35 & 91.53 & 97.18 & 98.73 & 99.31 & 91.62 \\
            & & 72.57 & 85.44 & 92.20 & 94.98 & 96.44 & & 88.33\\
            Affine & uniform (\textit{sampled}) & 67.52 & 91.13 & 97.01 & 98.64 & 99.26 & 90.71 \\
            & & 72.73 & 85.47 & 91.90 & 93.90 & 90.49 & & 86.78 \\
            Affine (class) & uniform (\textit{sampled}) & 66.97 & 90.99 & 97.06 & 98.69 & 99.28 & 90.60 \\
            & & 71.60 & 85.29 & 92.01 & 94.84 & 96.29 & & 88.01 \\
            Affine (full) & uniform (\textit{sampled}) & 66.08 & 90.37 & 96.82 & 98.51 & 99.14 & 90.18 \\
            & & 72.46 & 84.87 & 91.57 & 94.41 & 95.91 & & 87.84 \\
            \bottomrule
        \end{tabular}
        \caption{Continuation of table \ref{tab:result_hepimb}.}
        \label{tab:result_hepimb2}
    \end{table}
\endgroup

\newpage
\subsection{Discussion}
In our empirical comparison among network architectures, background distribution, and training procedure, we can conclude that:
\begin{enumerate}
    \item The affine-conditioning mechanism is able to better exploit the information brought by the mass feature, resulting in improved classification performance.
    \item The balanced training procedure, that yields balanced mini-batches, can further improve performance, even without changing the network architecture.
    \item The way the mass feature is distributed has a profound impact on how the model classifies and interpolates the missing masses. In general, the uniform distribution tends to easily overfit resulting in lower performance.
    \item Finally, the right combination of network architecture, background distribution, and balanced training, allowed us to greatly improve on both imbalanced classification and interpolation, almost recovering the classification performances achieved on the original, full, and not-imbalanced dataset.
\end{enumerate}

\newpage

\section{Conclusions}
\label{sec:conclusions}

In this study we first discussed the concept of "parametrization" which is really a re-brand of the widely used conditioning mechanisms in deep learning. Establishing such connection allows us to brought ideas and methods from such area, to improve pNNs in HEP. Another proposed intuition is about the structure of the data we use for signal-background classification: we know the contribution of the background(s), and at which mass the signal is generated. In fact we leverage the latter information to build a mass feature that parametrizes a neural network, allowing the model to replace a set of individual classifiers, as well as to interpolate beyond events seen during training. By studying the general structure of the data, we can exploit the inductive biases it provides by embedding them in the network design, and training as well. Lastly we demonstrated that pNNs are able to interpolate under real-world assumptions. We hope the ideas proposed here to be inspirational for further work about parametric networks, but also to be useful in other fields beyond HEP that have a similar problem setting and requirements.


\paragraph{Open Questions.} Our work is a first step towards a full understanding of parametric networks. We believe more properties and extensions to what is presented here to exist. In particular, we may suggest further research directions:
\begin{itemize}
    \item Real-world datasets are imbalanced, so either \textit{self-supervised learning}, class- or mass-specific \textit{data augmentation}, or (parametric) \textit{generative models} may provide a major improvement in classification performance.
    \item The signal is generated at few discrete mass hypotheses, what about parametrizing on the whole, \textit{continuous} mass range?
    \item The output of a pNN is a single number, why not letting the network output or discover a \textit{classification rule} that can be easily interpreted by physicists to further increase their knowledge about a certain phenomena?
\end{itemize}

\section*{Data Availability Statement}
The data that support the findings of this study are openly available at the following URL/DOI:
\url{https://zenodo.org/record/6453048}. The code used to produce our experiments is openly available on
GitHub: \url{https://github.com/Luca96/affine-parametric-networks}.

\section*{Acknowledgments}
The authors gratefully acknowledge the CMS Bologna analysis team - in particular Federica Primavera,
Stefano Marcellini and Gianni Masetti - for the constructive discussions, and Andrea Perrotta for the
valuable feedback and support.

\clearpage

\bibliographystyle{ieeetr}  
\bibliography{main}

\end{document}